\documentclass[aps,prd,twocolumn,superscriptaddress]{revtex4-1}
\usepackage{graphicx,amssymb,amsmath,times,color}

\begin{document}


\title{Search for transient gravitational waves in coincidence with short-duration radio transients during 2007--2013}

\author{%
B.~P.~Abbott,$^{1}$  
R.~Abbott,$^{1}$  
T.~D.~Abbott,$^{2}$  
M.~R.~Abernathy,$^{1}$  
F.~Acernese,$^{3,4}$ 
K.~Ackley,$^{5}$  
C.~Adams,$^{6}$  
T.~Adams,$^{7}$ 
P.~Addesso,$^{8}$  
R.~X.~Adhikari,$^{1}$
% ...
\\
\medskip
(LIGO Scientific Collaboration and Virgo Collaboration) 
\\
AND
\\
\medskip
A.~M.~Archibald,$^{136}$
S.~Banaszak,$^{17}$
A.~Berndsen,$^{137}$
J.~Boyles,$^{138}$
R.~F.~Cardoso,$^{139}$
P.~Chawla,$^{140}$
A.~Cherry,$^{137}$
L.~P.~Dartez,$^{86}$
D.~Day,$^{17}$
C.~R.~Epstein,$^{141}$
A.~J.~Ford,$^{86}$
J.~Flanigan,$^{17}$
A.~Garcia,$^{86}$
J.~W.~T.~Hessels,$^{136,142}$
J.~Hinojosa,$^{86}$
F.~A.~Jenet,$^{86}$
C.~Karako-Argaman,$^{140}$
V.~M.~Kaspi,$^{140}$
E.~F.~Keane,$^{143}$
V.~I.~Kondratiev,$^{136,144}$
M.~Kramer,$^{145,146}$
S.~Leake,$^{86}$
D.~Lorimer,$^{139}$
G.~Lunsford,$^{86}$
R.~S.~Lynch,$^{147}$
J.~G.~Martinez,$^{86}$
A.~Mata,$^{86}$
M.~A.~McLaughlin,$^{139}$
C.~A.~McPhee,$^{137}$
T.~Penucci,$^{148}$
S.~Ransom,$^{149}$
M.~S.~E.~Roberts,$^{150}$
M.~D.~W.~Rohr,$^{17}$
I.~H.~Stairs,$^{137}$
K.~Stovall,$^{151}$
J.~van Leeuwen,$^{136,142}$
A.~N.~Walker,$^{17}$
and
B.~L.~Wells$^{17,152}$%
\\
\medskip
{{}$^{\dag}$Deceased, May 2015. {}$^{\ddag}$Deceased, March 2015. }%
}\noaffiliation
\affiliation {LIGO, California Institute of Technology, Pasadena, CA 91125, USA }
\affiliation {Louisiana State University, Baton Rouge, LA 70803, USA }
\affiliation {Universit\`a di Salerno, Fisciano, I-84084 Salerno, Italy }
\affiliation {INFN, Sezione di Napoli, Complesso Universitario di Monte S.Angelo, I-80126 Napoli, Italy }
\affiliation {University of Florida, Gainesville, FL 32611, USA }
\affiliation {LIGO Livingston Observatory, Livingston, LA 70754, USA }
\affiliation {Laboratoire d'Annecy-le-Vieux de Physique des Particules (LAPP), Universit\'e Savoie Mont Blanc, CNRS/IN2P3, F-74941 Annecy-le-Vieux, France }
\affiliation {University of Sannio at Benevento, I-82100 Benevento, Italy and INFN, Sezione di Napoli, I-80100 Napoli, Italy }
\affiliation {Albert-Einstein-Institut, Max-Planck-Institut f\"ur Gravi\-ta\-tions\-physik, D-30167 Hannover, Germany }
\affiliation {Nikhef, Science Park, 1098 XG Amsterdam, The Netherlands }
\affiliation {LIGO, Massachusetts Institute of Technology, Cambridge, MA 02139, USA }
\affiliation {Instituto Nacional de Pesquisas Espaciais, 12227-010 S\~{a}o Jos\'{e} dos Campos,S\~{a}o Paulo, Brazil }
\affiliation {INFN, Gran Sasso Science Institute, I-67100 L'Aquila, Italy }
\affiliation {INFN, Sezione di Roma Tor Vergata, I-00133 Roma, Italy }
\affiliation {Inter-University Centre for Astronomy and Astrophysics, Pune 411007, India }
\affiliation {International Centre for Theoretical Sciences, Tata Institute of Fundamental Research, Bangalore 560012, India }
\affiliation {University of Wisconsin-Milwaukee, Milwaukee, WI 53201, USA }
\affiliation {Leibniz Universit\"at Hannover, D-30167 Hannover, Germany }
\affiliation {Universit\`a di Pisa, I-56127 Pisa, Italy }
\affiliation {INFN, Sezione di Pisa, I-56127 Pisa, Italy }
\affiliation {Australian National University, Canberra, Australian Capital Territory 0200, Australia }
\affiliation {The University of Mississippi, University, MS 38677, USA }
\affiliation {California State University Fullerton, Fullerton, CA 92831, USA }
\affiliation {LAL, Univ. Paris-Sud, CNRS/IN2P3, Universit\'e Paris-Saclay, Orsay, France }
\affiliation {Chennai Mathematical Institute, Chennai 603103, India }
\affiliation {Universit\`a di Roma Tor Vergata, I-00133 Roma, Italy }
\affiliation {University of Southampton, Southampton SO17 1BJ, United Kingdom }
\affiliation {Universit\"at Hamburg, D-22761 Hamburg, Germany }
\affiliation {INFN, Sezione di Roma, I-00185 Roma, Italy }
\affiliation {Albert-Einstein-Institut, Max-Planck-Institut f\"ur Gravitations\-physik, D-14476 Potsdam-Golm, Germany }
\affiliation {APC, AstroParticule et Cosmologie, Universit\'e Paris Diderot, CNRS/IN2P3, CEA/Irfu, Observatoire de Paris, Sorbonne Paris Cit\'e, F-75205 Paris Cedex 13, France }
\affiliation {Montana State University, Bozeman, MT 59717, USA }
\affiliation {Universit\`a di Perugia, I-06123 Perugia, Italy }
\affiliation {INFN, Sezione di Perugia, I-06123 Perugia, Italy }
\affiliation {European Gravitational Observatory (EGO), I-56021 Cascina, Pisa, Italy }
\affiliation {Syracuse University, Syracuse, NY 13244, USA }
\affiliation {SUPA, University of Glasgow, Glasgow G12 8QQ, United Kingdom }
\affiliation {LIGO Hanford Observatory, Richland, WA 99352, USA }
\affiliation {Wigner RCP, RMKI, H-1121 Budapest, Konkoly Thege Mikl\'os \'ut 29-33, Hungary }
\affiliation {Columbia University, New York, NY 10027, USA }
\affiliation {Stanford University, Stanford, CA 94305, USA }
\affiliation {Universit\`a di Padova, Dipartimento di Fisica e Astronomia, I-35131 Padova, Italy }
\affiliation {INFN, Sezione di Padova, I-35131 Padova, Italy }
\affiliation {CAMK-PAN, 00-716 Warsaw, Poland }
\affiliation {University of Birmingham, Birmingham B15 2TT, United Kingdom }
\affiliation {Universit\`a degli Studi di Genova, I-16146 Genova, Italy }
\affiliation {INFN, Sezione di Genova, I-16146 Genova, Italy }
\affiliation {RRCAT, Indore MP 452013, India }
\affiliation {Faculty of Physics, Lomonosov Moscow State University, Moscow 119991, Russia }
\affiliation {SUPA, University of the West of Scotland, Paisley PA1 2BE, United Kingdom }
\affiliation {University of Western Australia, Crawley, Western Australia 6009, Australia }
\affiliation {Department of Astrophysics/IMAPP, Radboud University Nijmegen, P.O. Box 9010, 6500 GL Nijmegen, The Netherlands }
\affiliation {Artemis, Universit\'e C\^ote d'Azur, CNRS, Observatoire C\^ote d'Azur, CS 34229, Nice cedex 4, France }
\affiliation {MTA E\"otv\"os University, ``Lendulet'' Astrophysics Research Group, Budapest 1117, Hungary }
\affiliation {Institut de Physique de Rennes, CNRS, Universit\'e de Rennes 1, F-35042 Rennes, France }
\affiliation {Washington State University, Pullman, WA 99164, USA }
\affiliation {Universit\`a degli Studi di Urbino 'Carlo Bo', I-61029 Urbino, Italy }
\affiliation {INFN, Sezione di Firenze, I-50019 Sesto Fiorentino, Firenze, Italy }
\affiliation {University of Oregon, Eugene, OR 97403, USA }
\affiliation {Laboratoire Kastler Brossel, UPMC-Sorbonne Universit\'es, CNRS, ENS-PSL Research University, Coll\`ege de France, F-75005 Paris, France }
\affiliation {Astronomical Observatory Warsaw University, 00-478 Warsaw, Poland }
\affiliation {VU University Amsterdam, 1081 HV Amsterdam, The Netherlands }
\affiliation {University of Maryland, College Park, MD 20742, USA }
\affiliation {Center for Relativistic Astrophysics and School of Physics, Georgia Institute of Technology, Atlanta, GA 30332, USA }
\affiliation {Laboratoire des Mat\'eriaux Avanc\'es (LMA), CNRS/IN2P3, F-69622 Villeurbanne, France }
\affiliation {Universit\'e Claude Bernard Lyon 1, F-69622 Villeurbanne, France }
\affiliation {Universitat de les Illes Balears, IAC3---IEEC, E-07122 Palma de Mallorca, Spain }
\affiliation {Universit\`a di Napoli 'Federico II', Complesso Universitario di Monte S.Angelo, I-80126 Napoli, Italy }
\affiliation {NASA/Goddard Space Flight Center, Greenbelt, MD 20771, USA }
\affiliation {Canadian Institute for Theoretical Astrophysics, University of Toronto, Toronto, Ontario M5S 3H8, Canada }
\affiliation {Tsinghua University, Beijing 100084, China }
\affiliation {University of Michigan, Ann Arbor, MI 48109, USA }
\affiliation {National Tsing Hua University, Hsinchu City, 30013 Taiwan, Republic of China }
\affiliation {Charles Sturt University, Wagga Wagga, New South Wales 2678, Australia }
\affiliation {University of Chicago, Chicago, IL 60637, USA }
\affiliation {Caltech CaRT, Pasadena, CA 91125, USA }
\affiliation {Korea Institute of Science and Technology Information, Daejeon 305-806, Korea }
\affiliation {Carleton College, Northfield, MN 55057, USA }
\affiliation {Universit\`a di Roma 'La Sapienza', I-00185 Roma, Italy }
\affiliation {University of Brussels, Brussels 1050, Belgium }
\affiliation {Sonoma State University, Rohnert Park, CA 94928, USA }
\affiliation {Texas Tech University, Lubbock, TX 79409, USA }
\affiliation {Cardiff University, Cardiff CF24 3AA, United Kingdom }
\affiliation {University of Minnesota, Minneapolis, MN 55455, USA }
\affiliation {The University of Melbourne, Parkville, Victoria 3010, Australia }
\affiliation {The University of Texas Rio Grande Valley, Brownsville, TX 78520, USA }
\affiliation {The University of Sheffield, Sheffield S10 2TN, United Kingdom }
\affiliation {Universit\`a di Trento, Dipartimento di Fisica, I-38123 Povo, Trento, Italy }
\affiliation {INFN, Trento Institute for Fundamental Physics and Applications, I-38123 Povo, Trento, Italy }
\affiliation {Montclair State University, Montclair, NJ 07043, USA }
\affiliation {The Pennsylvania State University, University Park, PA 16802, USA }
\affiliation {National Astronomical Observatory of Japan, 2-21-1 Osawa, Mitaka, Tokyo 181-8588, Japan }
\affiliation {School of Mathematics, University of Edinburgh, Edinburgh EH9 3FD, United Kingdom }
\affiliation {Indian Institute of Technology, Gandhinagar Ahmedabad Gujarat 382424, India }
\affiliation {Institute for Plasma Research, Bhat, Gandhinagar 382428, India }
\affiliation {University of Szeged, D\'om t\'er 9, Szeged 6720, Hungary }
\affiliation {Embry-Riddle Aeronautical University, Prescott, AZ 86301, USA }
\affiliation {Tata Institute of Fundamental Research, Mumbai 400005, India }
\affiliation {INAF, Osservatorio Astronomico di Capodimonte, I-80131, Napoli, Italy }
\affiliation {American University, Washington, D.C. 20016, USA }
\affiliation {University of Massachusetts-Amherst, Amherst, MA 01003, USA }
\affiliation {University of Adelaide, Adelaide, South Australia 5005, Australia }
\affiliation {West Virginia University, Morgantown, WV 26506, USA }
\affiliation {University of Bia{\l }ystok, 15-424 Bia{\l }ystok, Poland }
\affiliation {SUPA, University of Strathclyde, Glasgow G1 1XQ, United Kingdom }
\affiliation {IISER-TVM, CET Campus, Trivandrum Kerala 695016, India }
\affiliation {Northwestern University, Evanston, IL 60208, USA }
\affiliation {Institute of Applied Physics, Nizhny Novgorod, 603950, Russia }
\affiliation {Pusan National University, Busan 609-735, Korea }
\affiliation {Hanyang University, Seoul 133-791, Korea }
\affiliation {NCBJ, 05-400 \'Swierk-Otwock, Poland }
\affiliation {IM-PAN, 00-956 Warsaw, Poland }
\affiliation {Rochester Institute of Technology, Rochester, NY 14623, USA }
\affiliation {Monash University, Victoria 3800, Australia }
\affiliation {Seoul National University, Seoul 151-742, Korea }
\affiliation {University of Alabama in Huntsville, Huntsville, AL 35899, USA }
\affiliation {ESPCI, CNRS, F-75005 Paris, France }
\affiliation {Universit\`a di Camerino, Dipartimento di Fisica, I-62032 Camerino, Italy }
\affiliation {Southern University and A\&M College, Baton Rouge, LA 70813, USA }
\affiliation {College of William and Mary, Williamsburg, VA 23187, USA }
\affiliation {Instituto de F\'\i sica Te\'orica, University Estadual Paulista/ICTP South American Institute for Fundamental Research, S\~ao Paulo SP 01140-070, Brazil }
\affiliation {University of Cambridge, Cambridge CB2 1TN, United Kingdom }
\affiliation {IISER-Kolkata, Mohanpur, West Bengal 741252, India }
\affiliation {Rutherford Appleton Laboratory, HSIC, Chilton, Didcot, Oxon OX11 0QX, United Kingdom }
\affiliation {Whitman College, 345 Boyer Avenue, Walla Walla, WA 99362 USA }
\affiliation {National Institute for Mathematical Sciences, Daejeon 305-390, Korea }
\affiliation {Universit\'e de Lyon, F-69361 Lyon, France }
\affiliation {Hobart and William Smith Colleges, Geneva, NY 14456, USA }
\affiliation {Janusz Gil Institute of Astronomy, University of Zielona G\'ora, 65-265 Zielona G\'ora, Poland }
\affiliation {Andrews University, Berrien Springs, MI 49104, USA }
\affiliation {Universit\`a di Siena, I-53100 Siena, Italy }
\affiliation {Trinity University, San Antonio, TX 78212, USA }
\affiliation {University of Washington, Seattle, WA 98195, USA }
\affiliation {Kenyon College, Gambier, OH 43022, USA }
\affiliation {Abilene Christian University, Abilene, TX 79699, USA }
\affiliation {ASTRON, the Netherlands Institute for Radio Astronomy, Postbus 2, 7990 AA Dwingeloo, The Netherlands}
\affiliation {Department of Physics and Astronomy, University of British Columbia, 6224 Agricultural Road, Vancouver, British Columbia V6T 1Z1, Canada}
\affiliation {Department of Physics and Astronomy, Western Kentucky University, Bowling Green, KY 42101, USA}
\affiliation {Dept. of Physics and Astronomy, West Virginia University, Morgantown, WV 26506, USA}
\affiliation {Department of Physics \& McGill Space Institute, McGill University, 3600 Univesity Street, Montreal, QC H3A 2T8, Canada}
\affiliation {Department of Astronomy, Ohio State University, 140 W. 18th Avenue, Columbus, OH 43210, USA}
\affiliation {Anton Pannekoek Institute for Astronomy, University of Amsterdam, Science Park 904, 1098 XH Amsterdam, The Netherlands}
\affiliation {SKA Organization, Jodrell Bank Observatory, SK11 9DL, UK}
\affiliation {Astro Space Center of the Lebedev Physical Institute, Profsoyuznaya str. 84/32, Moscow 117997, Russia}
\affiliation {Max Planck Institute for Radio Astronomy, Auf dem Hugel 69, 53121 Bonn, Germany}
\affiliation {The University of Manchester, Oxford Rd, Manchester M13 9PL, UK}
\affiliation {National Radio Astronomy Observatory, Green Bank, WV 24944, US}
\affiliation {Department of Astronomy, University of Virginia, Charlottesville, VA 22904-4325, USA}
\affiliation {National Radio Astronomy Observatory, Charlottesville, VA 22903, USA}
\affiliation {Eureka Scientific, Inc., 2452 Delmer Street, Suite 100, Oakland, CA 94602-3017, USA}
\affiliation {Department of Physics and Astronomy, University of New Mexico, Albuquerque, NM, USA}
\affiliation {Department of Atmospheric Science, Colorado State University, Fort Collins, CO, USA}

\noaffiliation

\begin{abstract}

We present an archival search for transient gravitational-wave bursts in coincidence
with 27 single-pulse triggers from Green Bank Telescope pulsar surveys, using
the LIGO, Virgo and GEO interferometer network. We also discuss a check
for gravitational-wave signals in coincidence with Parkes fast radio
bursts using similar methods.  Data analyzed in these searches were collected
between 2007 and 2013.
Possible sources of emission of both short-duration radio signals and
transient gravitational-wave emission include starquakes on neutron stars,
binary coalescence of neutron stars, and cosmic string cusps.  While no
evidence for gravitational-wave emission in coincidence with these radio
transients was found, the current analysis serves as a prototype for
similar future searches using more sensitive second-generation interferometers.

\end{abstract}

\maketitle

\section{Introduction}

Plausible models for coincident or near-coincident emission of both radio and gravitational-wave (GW) transients exist
for a number of astrophysical phenomena, including single neutron stars,
merging neutron star binaries and cosmic string cusps.
Identification of a GW in close temporal and spatial
coincidence with a fast radio burst or other radio transient could place
significant constraints on the source of emission, with further constraints possible
based on the morphology of the gravitational-wave signal.
In this paper we present a search for GWs in coincidence
 with millisecond-scale duration radio transient pulses.
We have conducted externally
triggered searches for gravitational waves with the LIGO-Virgo-GEO network in
coincidence with both Galactic single-pulse pulsar candidates from the Green
Bank Telescope and a sample of cosmological fast radio burst (FRB) candidates from the Parkes
Telescope.  Individual radio pulses range from 1 to tens of ms in duration and
were observed in frequency bands from hundreds of MHz to 1 GHz.  The LIGO Scientific Collaboration (LSC) and Virgo Collaboration regularly search for continuous gravitational
waves arriving from the direction of known radio pulsars (see, \emph{e.g.} Refs. \cite{pulsar1,pulsar2} for recent examples) and a search for
gravitational waves in coincidence with a Vela pulsar glitch was conducted
previously \cite{clark}.

The present work marks the first LIGO-Virgo search in coincidence
with radio transients and serves as a prototype for searches with advanced
interferometers.  Given that the origin of these radio
transients is currently unclear, our analysis is designed to search broadly
for a gravitational-wave transient burst, without requiring a specific
type of waveform.

The paper is organized as follows: after briefly describing the network of
gravitational-wave interferometers used in this analysis in section II,
we discuss possible mechanisms leading to joint emission of $\sim$few hundred
hertz gravitational waves and radio transient signals.  We describe the radio data
 used in the analysis in  section IV, followed by the gravitational-wave
search methods and results in sections V and VI, respectively.  We conclude
with a discussion of future prospects for joint analysis of radio and
gravitational-wave data in section VII.

\section{Gravitational-Wave Interferometer Network}

The LIGO Scientific Collaboration and Virgo Collaboration operate a network of
power-recycled Fabry-Perot Michelson interferometers designed to be sensitive to very small
relative changes in length (on the order of one part in 10$^{21}$) of the two
orthogonal detector arms.  LIGO operates two sites in the United States, one
in Livingston Parish, Louisiana and another at the Hanford site
in Washington.  Both LIGO facilities operate an interferometer with an arm length of 4 km 
(called L1 and H1, respectively) and Hanford operated an additional 
smaller, collocated interferometer (H2) until September 2007 \cite{LIGO}.
The LIGO Scientific Collaboration also operates a 600 m interferometer, GEO\,600, 
near Hannover, Germany (G1) \cite{GEO}.  The Virgo Collaboration operates a single
3 km interferometer near Cascina, Italy (V1) \cite{Virgo}.

Since this paper involves the analysis of radio transients across a period of several
years of initial detector data, multiple science runs of these interferometers are used.  Data analyzed in
this paper are drawn from summer 2007, coincident with LIGO's fifth and Virgo's first science
run, as well as late 2009, coincident with LIGO's sixth and Virgo's third science run.
FRB candidates discussed in this paper are coincident with GEO\,600 Astrowatch 
data ranging from 2011 to 2013, and in some cases Virgo's fourth science run, which took
place in summer 2011. See \cite{geogrb} for a comparison of sensitivities for
these instruments from 2007 to 2014.

The LIGO-Virgo network has undergone extensive upgrades to second-generation instruments, and during the first Advanced LIGO observation run
made the first direct detection of a gravitational-wave transient \cite{discovery}.  After reaching design sensitivity the Advanced LIGO \cite{aLIGO} and Advanced Virgo \cite{aVirgo}
detectors will have an order of magnitude improvement in range relative
to their first-generation counterparts.  
Additional advanced interferometers are scheduled to join
the global network in the future, including Kagra in Japan \cite{Kagra} and a third LIGO site
in India \cite{LIGOIndia}.

\section{Potential Sources of Joint Radio and Gravitational-Wave Emission}

There are a number of astrophysical phenomena that may plausibly produce 
gravitational waves in close coincidence with radio frequency emission. We 
focus this discussion on a few types of sources which may produce both GWs
and radio pulses with frequency and duration suitable to the instruments
being used in this analysis.  More detailed discussion can be found in \cite{predoi}.

\subsection{Single neutron stars}

Transient gravitational-wave emission can occur when a temporary deformation
of a rapidly rotating neutron star creates a quadrupolar moment.  Typically,
this is believed to happen as a result of crust cracking from magnetic, 
gravitational or superfluid forces, dubbed a starquake \cite{starquakes,duncan}, or
from other asteroseismic phenomena resulting in the shifting of the neutron star's
 crust \cite{kokkotas}.  Asteroseismology may result in several
types of quasinormal oscillatory modes of the neutron star which could produce GW emission.  
These include torsional modes at low frequencies \cite{zink} and the f-mode, with GW emission 
believed to typically peak around 2 kHz \cite{fmode}.
The amplitude of the GW emission even in optimistic cases, however, is small
enough that sensitivity to this type of source will be limited to our own
Galaxy even in the advanced detector era.

Radio pulsars result from beamed emission from the poles of a rapidly 
rotating, highly magnetized neutron star sweeping past the Earth, producing reliably periodic radio signals.  The asteroseismic events described 
above may result in a distinct increase in the rotation rates of these neutron
stars, typically followed by a gradual return to their original period.  This
phenomenon, called a pulsar glitch,  has been observed across a large number
 of pulsars, especially younger ones (see \emph{e.g.} Ref. \cite{pulsarglitch} and 
references therein). A search for gravitational-wave emission from quasinormal
modes in coincidence with the observed glitching of pulsars was the subject of
a previous LSC publication \cite{clark}. Models for neutron star 
asteroseismic phenomena similar to those under discussion have also motivated 
previous gravitational-wave searches in coincidence with soft gamma repeater flares \cite{SGR}.

A related phenomenon to radio pulsars
 is the rotating radio transient (RRAT).  RRATs emit short-duration 
radio pulses similar in character to pulsars, but are distinguished 
by their lack of predictable periodic behavior.  RRATs may be ``dying'' pulsars near the end of their life cycles,
neutron stars with especially high magnetization, or conventional pulsars 
of which the observation is often obscured by intervening matter between the pulsar 
and Earth, although it is also possible that other phenomena may manifest 
observationally as RRATs \cite{RRATS}.

The standard indication of an asteroseismic event in an isolated neutron star
is a pulsar glitch, but there are plausible mechanisms that could result
in the observation of a transient radio pulse.  This could simply be through
the pulsar radio emission coming into view from the Earth as the pulsar's orbit
shifts slightly, but there is also some evidence that pulsarlike radio emission
can be ``switched on'' in coincidence with a glitching mechanism 
\cite{keane,lyne,weltevrede}.  For some models, gravitational waves emitted
by neutron stars are predicted to be 
detectable at a distance scale on the order of kiloparsecs with first generation of
interferometers.  We therefore consider single neutron stars as 
possible sources of coincident GW and radio transient events.  

\subsection{Binary neutron star coalescence}

The most easily observable transient gravitational-wave signature in the 
frequency range of LIGO and Virgo is the merger of a binary system of compact
objects, specifically neutron stars or black holes.  
In the final moments before
 the compact objects merge, the upward sweep in frequency of the gravitational
 wave emission is predicted to produce a characteristic chirplike signal. 
Recent evidence suggests that neutron star binary mergers may create at least some
fraction of FRBs \cite{FRBhost}.

Compact binary coalescence is currently the only confirmed source of 
directly detectable gravitational waves \cite{discovery}.
Once design sensitivity is reached in 
the advanced detector era, the ground-based network of interferometers is
predicted to detect several to a few hundred binary coalescence GW
signals per year of operation \cite{cbcrates}.

There are several models for radio emission in coincidence with a compact
binary coalescence GW signal.  This may be pulsarlike radio emission, either
from the reactivation of the dormant pulsar emission in one of the neutron
stars through interactions prior to merger \cite{lipunov} or by a 
hypermassive neutron star, which may sometimes be produced as an intermediate state
before collapse to a black hole \cite{pshirkov}.  Another possible 
mechanism is the radiation at radio frequencies as a result of magnetospheric 
interactions \cite{hansen}.

Given an appropriate density in the surrounding environment, the gravitational
waves emitted by a compact binary coalescence may induce electromagnetic
radiation through magnetohydrodynamic interactions.  While this interaction
would directly produce radiation at the same relatively low frequencies as
the GWs themselves, upconversion through inverse Compton radiation may result
in emission at radio frequencies \cite{moortgat}.  This particular 
magnetohydrodynamic mechanism does not necessarily require neutron star 
coalescence as the mechanism for production of the GWs, but this class of
source is likely to be able to produce GWs of suitable amplitude and may
be surrounded by an environment suitable to this mechanism \cite{moortgat2}.

While the sensitivity of the LIGO/Virgo network to gravitational waves from compact 
binary coalescence scenarios is dependent on the mass, spin, and other properties of the merging objects, interferometers in the initial detector era were
typically sensitive to mergers of two neutron stars out to a distance on the
order of 10 Mpcs \cite{cbcrates}.

\subsection{Cosmic strings}
 
Cosmic strings, formed during symmetry breaking in the early Universe, are
topological defects thought to be capable of emitting large amounts of energy
from their cusps or kinks \cite{berezinsky}.
A cosmic string cusp may emit gravitational waves with a $f^{-4/3}$ 
frequency dependence up to a cutoff frequency \cite{damour}, potentially at 
frequencies and amplitudes detectable by ground-based interferometers 
\cite{siemens,cssearch}.  The same cusps may produce short-duration linearly
 polarized radio bursts \cite{cai}, a mechanism that has previously been
proposed as the origin of the original Lorimer burst \cite{vachaspati}.
Unlike GWs from other sources discussed, cosmic
string cusps could theoretically produce detectable GWs at cosmological
distances, which makes them particularly interesting in the context of Parkes
FRBs with dispersion measures (DMs) indicative of cosmological distances.

\subsection{Other potential sources}

The three classes of sources resulting in simultaneous GW and radio emission
described above are not an exhaustive list of theoretical joint sources, but
most other types of sources are outside the scope of the analyses described in
this paper due to the frequency or duration of the predicted GW and/or radio
emission not being well suited to the instruments described in this analysis.
For example, some scenarios in which gamma-ray bursts (GRBs) may also result in
radio emission are not explicitly considered in developing this analysis;
prompt radio emission models \cite{usov,sagiv}
predict signals at much lower frequencies than the Green Bank and Parkes 
telescopes can detect, and GRB radio afterglows \cite{nakar} occur on 
longer time scales inconsistent with the short radio pulses that are 
the subject of the searches described in this paper.  Core-collapse supernovae have also
been proposed as plausible sources of short-duration radio pulses \cite{keane} and GW emission.
However, we do not explicitly include supernovae among the classes of emission for which we are searching 
when designing the analysis as there are no observed nearby core-collapse supernovae in close coincidence
with the radio transients under consideration.

\section{Radio Pulse Data}

\subsection{Green Bank single pulse analysis data}

The Robert C. Byrd Green Bank Telescope is the world's largest fully steerable 
single-dish radio telescope.  In the summer of 2007 a drift-scan pulsar survey was
conducted in a band of 350$\pm$25 MHz \cite{gbtdriftscan}.  This time frame was
during Initial LIGO's fifth science run and Initial Virgo's first science run.
In addition to the 
identification of continuously
observable pulsars, a ``single-pulse'' archival search was performed to look
for transient emission of millisecond-scale duration radio pulses.  The drift-scan
team provided LIGO/Virgo with 33 of these observed single-pulse triggers, ten of
which were confirmed to originate from sources with repeated emission through
followup observations, thus most likely originating from a pulsar or 
RRAT.  Some of the triggers exhibited only a single radio pulse, while others
show several pulses within a 2 min window, but in order to be considered a 
viable astrophysical signal, all pulses were required to exhibit the $1/f^{2}$
dispersion behavior expected as a result of dispersion in the interstellar 
medium.

For each of the 33 candidates, right ascension, declination, dispersion measure,
and arrival time at solar system barycenter were provided.  
The dispersion measures provided (between 15 and 170 pc cm$^{-3}$) are in 
general consistent with a population of sources from within our own Galaxy.
For purposes of the gravitational-wave search, barycentric arrival times were 
adjusted to UTC arrival times at the detector using code previously applied to
 LIGO pulsar analyses \cite{pulsar1} and cross-checked against conversions to 
the detector frame provided by Green Bank for a subset of triggers.  

A survey of the Galaxy's Northern Celestial Cap was conducted with the
Green Bank Telescope in 2009 and 2010 \cite{GBNCC}.  The single-pulse analysis
searching for RRATs or related phenomena was more automated than in the
drift-scan analysis and resulted in seven published candidates being reported.
These radio triggers corresponded to Initial LIGO's sixth and Initial Virgo's
third science runs, and were treated identically to drift-scan triggers for
GW analysis purposes.

\subsection{Parkes fast radio bursts}

The report of FRBs originating from apparently
cosmological distances \cite{parkesfrbs,parkes131104,parkes140514}
has led to an increased interest
in short-duration radio transients.  These radio transients resemble the
original Lorimer burst \cite{lorimerburst} reported by Parkes in 2007.
Since then Arecibo and Green Bank have each reported an FRB
\cite{alfafrb,GBTFRB}.  
The origin of these FRBs is unclear, with several 
possible
astrophysical sources and terrestrial backgrounds posited as the origin of
these radio bursts.  Recent followup observations of the Arecibo burst in particular indicate it is a repeating phenomenon \cite{alfarepeat}, which substantially narrows the range of plausible sources.  An observed correlation with another astrophysical
signal would do much to clarify the nature of these bursts and help
confirm their nature as a previously unknown astrophysical phenomenon.

The FRBs of interest for a gravitational-wave search (i.e. omitting bursts
from 2001, prior to the construction of sensitive GW interferometers) have observed
fluences ranging from 0.55 to 7.3 Jy ms, observed width from 0.64 to 15.6 ms,
and dispersion measures suggesting cosmological origin, ranging from
563 to 1629 pc cm$^{-3}$ \cite{FRBCAT}.  
While most of these FRBs did not occur during LIGO/Virgo science runs, we 
performed a check in the Virgo science run data and GEO\,600 Astrowatch data when 
available.

Given the difference in the dispersion measure and other properties, the FRBs are most likely primarily a distinct
class of sources compared to the RRAT-like observations in the previous section.  Current evidence points to an astrophysical origin, especially as known perytons exhibit properties distinct from FRBs \cite{microperyton}.  Several of the emission
mechanisms discussed in Section III are considered viable candidate progenitors, including neutron star and
compact binary coalescing systems \cite{FRBCR,FRBgalactic,FRBBNS}. 

\subsection{Short-duration radio transients not analyzed}

Potential FRBs from sources other than the above were considered but were not 
coincident with an active network of GW interferometers. In some cases this was because they
occurred during times before sensitive GW data were available \cite{keaneburst,rubioherrera}. This includes
the  original Lorimer burst \cite{lorimerburst}. Neither radio transient reported to be
observed in coincidence with a GRB in Ref. \cite{bannister} was coincident with LIGO/Virgo data \cite{grbsearch}.
The first observation of Arecibo's reported FRB \cite{alfafrb,alfarepeat} occurred during the LIGO and Virgo network upgrade to advanced
interferometers and fell during a time in which GEO\,600 was not collecting science data as part of its Astrowatch 
program, with published followup observations also occurring prior to the first Advanced LIGO observation run.  Although FRB110523, identified by Green Bank Telescope \cite{GBTFRB}, occurred during Initial
Virgo's fourth science run, Virgo was not taking data at the time of the event.

\section{Analysis Method}

\subsection{Procedure}

The search for gravitational waves in coincidence with short radio transients
 was conducted using the {\sc X-Pipeline} analysis package \cite{xpipeline}.  
This software has been used for a number of GW searches in 
coincidence with astrophysical triggers. The analysis procedure for this
search was modeled directly after conceptually similar searches for GWs in
coincidence with gamma-ray bursts \cite{grbsearch,ipngrbsearch}, but the parameters 
were modified to account for the particular types of GW sources under consideration.
Similar adjustments between GRB and radio transient searches can be used to design
radio transient searches in advanced interferometers.

For each radio trigger analyzed, we use {\sc X-Pipeline} to conduct a coherent 
search for a GW signal consistent with the location and time of the
radio signal.  The physical locations of the individual GW interferometers in
the network and the antenna patterns based on their orientations are used to
reject potential GW signals that are not consistent with the radio
signal's sky location and only signals within $\pm$2 minutes of the radio
trigger are considered.  Coherent and incoherent energy combinations are
calculated for each potential trigger and a series of two-dimensional cuts
is applied to reject triggers physically inconsistent with a GW.  The
false alarm probability of any surviving triggers after all cuts is
estimated based on the ``time-lag'' method.  This utilizes interferometric
data outside but near the on-source window, but introduces hundreds of 
artificial time offsets between the interferometers that are much larger than 
the time of flight of a real GW signal in order to obtain statistics on the 
significance of background in which no coherent signal is present.  These
procedures are discussed in more detail in Ref. \cite{grbsearch}, with adjustments
for our specific analysis as described below.

\subsection{Analysis-specific search parameters}

Our frequency range, temporal and spatial coincidence windows, veto methods,
 and other 
parameters were selected to handle a range of possible astrophysical emission 
mechanisms consistent with short radio pulses as discussed in Section III.
 Where allowed by calibration \cite{calibrationS5}, the frequency range was 
between 64 
Hz and approximately 3 kHz.  The majority of GW searches cut off at 
2 kHz due to rising shot noise and increased computational costs
 at higher frequency. However, increasing the upper frequency range for this analysis
allows us to include a large subset of possible GW emission from
single neutron stars.
This increase in frequency also requires us to perform the search over a
much denser grid of points on the sky, but the excellent spatial resolution
of radio telescopes relative to many other astrophysical observations makes
this adjustment feasible.  

The on-source time window when searching for GW signals around the radio
pulse was taken to be $\pm 120$ s around the observed radio pulse.
While it is difficult to exhaustively cover all possible scenarios for
time separation between radio and GW emission, the time window selected
covers the offsets between emission for the range of 
scenarios informing the design of the analysis.
Since the radio pulse arrival times are corrected for dispersion
there is very little additional uncertainty in the time-of-flight
difference between the two types of emission.
For analyses in coincidence with Green Bank triggers, the angular uncertainty
on the sky is taken to be 0.55 deg.  This accounts for two effects,
including 95\% of Green Bank's total beam width for a 350 MHz signal and
including a small adjustment for the drift of the source across the sky 
during the time span over which {\sc X-Pipeline} conducts a test for a self-consistent
GW signal.

Unlike recent GW searches in coincidence with GRBs that used similar procedures \cite{ipngrbsearch}, 
our background vetoing procedures do not rely on the assumption that the 
GW signal will be circularly polarized.  While this is a reasonable assumption
for signals in coincidence with gamma-ray bursts, the greater variety of
possible astrophysical sources we consider in this analysis does not justify this assumption.

\subsection{Simulated waveforms}

While we do not require a particular signal morphology for our gravitational
wave signal, we tune the analysis and characterize its performance based on
an ensemble of simulated waveforms.  These ten waveforms 
include: 

\begin{itemize}
\item{damped sinusoids at peak frequencies of 1750 Hz and 2300 Hz and decay
time constants of 0.26 s and 0.13 s respectively,
representing typical emission expected from neutron star asteroseismic events
under different assumptions for the neutron star equation of state
 \cite{fmode}};
\item{cosmic string cusp waveforms with upper frequency limits of 300 Hz and 1000 Hz};
\item{linearly polarized gaussian-envelope sine waves at central frequencies of 235 and 945 Hz
with a quality factor of 9};
\item{circularly polarized gaussian-envelope sine waves at central frequencies of 150 Hz and 300 Hz, also with a quality factor of 9};
\item{a compact binary coalescence signal from merging 1.4 solar
mass neutron stars};
\item{a compact binary coalescence signal from a 1.4 solar mass neutron star
and a 50 solar mass black hole.} 
\end{itemize}

\noindent These waveforms were selected to broadly represent the types of gravitational
wave signals that may occur in coincidence with radio pulses without focusing
too heavily on a specific morphology.  In addition, the last three bullet 
points describe specific signals used in previous LIGO searches in order to
facilitate sensitivity comparisons with previous work.

\section{Analysis Results}

\subsection{Green Bank pulsar surveys}

Of the 33 single-pulse radio candidates from the Green Bank drift-scan survey,
 25 were analyzable with at least three interferometers in the LIGO-Virgo 
network.  Of the seven RRAT candidates identified in the Northern Celestial Cap
survey, only two were analyzable with two or more interferometers in the gravitational 
wave network.

None of these 27 radio pulses resulted in viable GW candidates.  The most significant
result for a single candidate was a 2.7$\%$ single trial false alarm probability 
for RRAT 1944-1017, which is completely consistent with 
background for an ensemble of 27 trials.  Table \ref{GBT} shows information about each
radio candidate, including information about the radio source, as well as 
GW network and upper limits on h$_\mathrm{rss}$ (root sum squared strain) for
three of the simulated GW waveforms.  The last two entries in the table are the Northern 
Celestial Cap survey triggers.

\begin{figure}
\hspace*{-0.5 cm}\mbox{
\includegraphics*[width=0.6\textwidth]{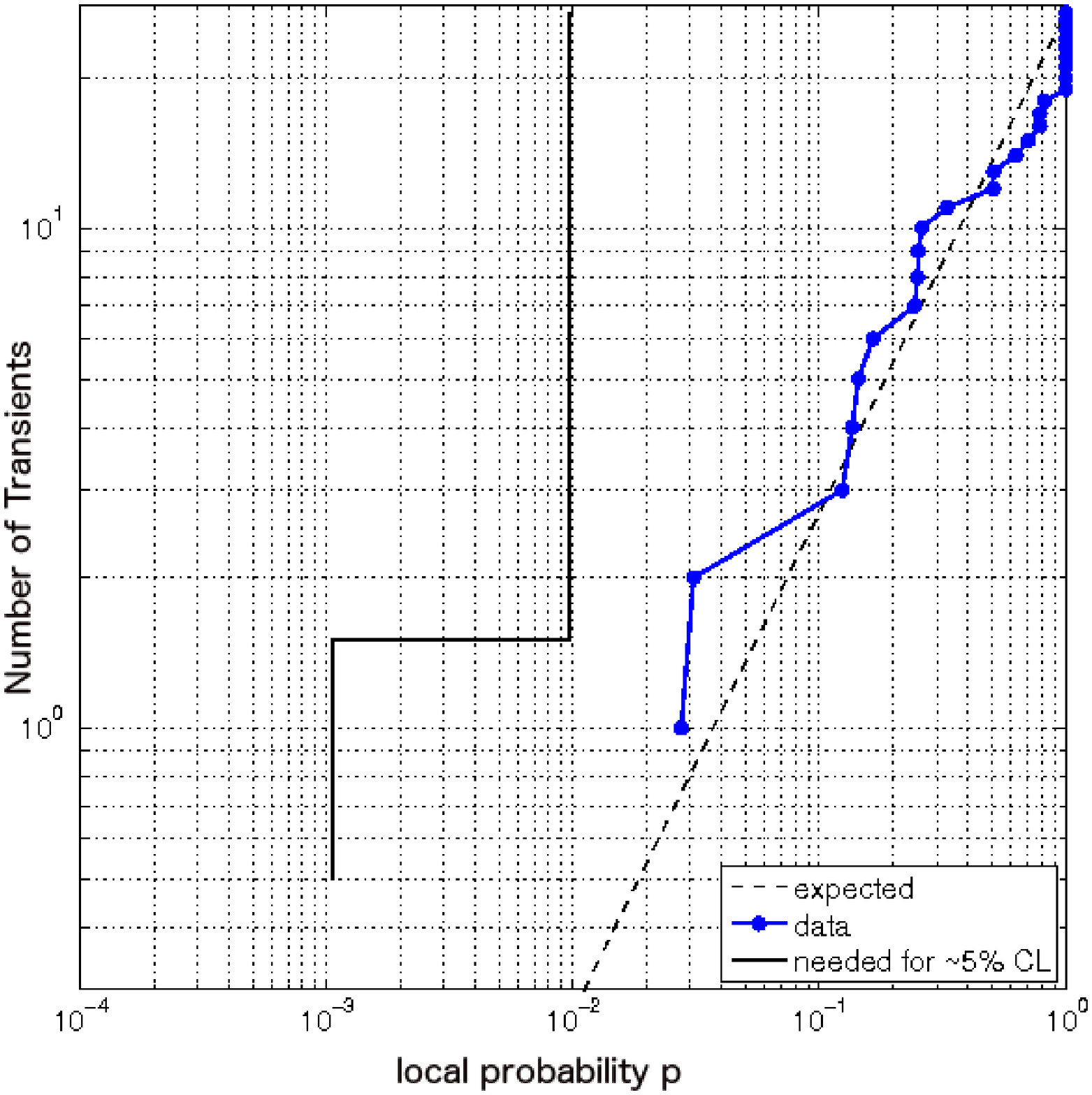}}
\caption{\label{binomial}Cumulative distribution of p-values from the analysis of 27 radio triggers from Green Bank for 
evidence of a GW transient associated with the event.  The expected distribution in the absence of a 
signal is indicated by the dashed line.  Points at p-value of unity are triggers with no event in the 
on-source region after selection cuts.}
\end{figure}

In addition to individual analysis of the radio candidates, we also perform a
weighted binomial test of the p-value distribution of the most significant
surviving trigger from the GW analysis, using the same methodology as employed
previously in searches for GWs in coincidence with GRBs \cite{grbsearch,geogrb}.
  This distribution is plotted against
expectation in Fig. \ref{binomial}.  The test yields a background probability
of 30\%, which is consistent with the null hypothesis.

\begin{figure}
\begin{center}
\hspace*{-0.5 cm}\mbox{
\includegraphics*[width=0.57\textwidth]{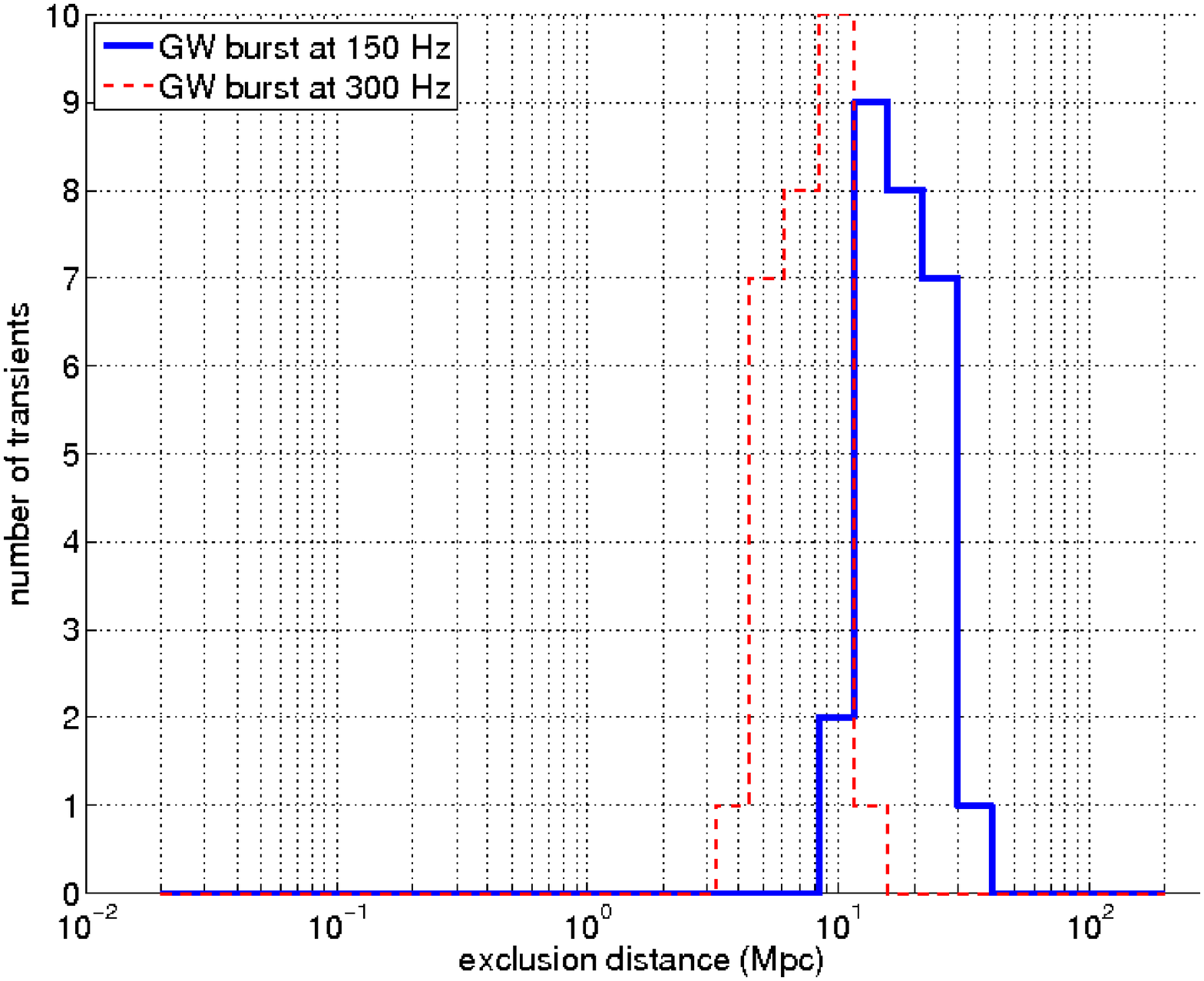}}
\caption{\label{dist_combined}Histograms for the sample of Green Bank radio transients of distance exclusions at 90\% confidence level for possible GRBs
associated with radio transients.  Waveforms are circularly polarized sine-Gaussian GW burst models with central frequency of 150 and 300 Hz.}
\end{center}
\end{figure}

Possible association between GRBs and FRBs has been widely discussed  
(see, {\it e.g.}, Refs. \cite{FRBhost,lipunov2, frbgrb, bannister}), with indications that at least a subset of
radio bursts may be coupled with gamma-ray bursts.  We therefore follow previous LIGO
analyses \cite{ipngrbsearch} and calculate 90\% confidence level exclusion distances,
for two of our simulated circularly polarized waveforms, assuming an optimistic 
standard siren in which $\sim$1\% of a solar mass is converted to gravitational-wave energy.  
A histogram of these distance constraints is 
shown in Fig. \ref{dist_combined}.  In general, limits in the few to tens of Mpc range indicate 
that we would be sensitive to a GW signal under these assumptions well outside of our own Galaxy, 
but at substantially less than the cosmological distances measured for FRBs.
For the 150 Hz sine-Gaussian waveform, using standard calculations \cite{GRB1}
about $\sim$4$\times$10$^{52}$ergs of
energy would have to be emitted for a detectable source emitting 
isotropically at a distance of 20 Mpc.

\subsection{Parkes Telescope FRBs}

We examined a list of 14 FRBs \cite{FRBCAT} from Parkes, occurring as early as 2001 but 
primarily concentrated within the last five years.  While none of the event times corresponded to science runs for 
Hanford or Livingston, eight of the FRBs corresponded to times when GEO\,600 Astrowatch data were available
and two of these also corresponded to data from Initial Virgo's fourth science run.  After omitting two of these
FRBs for which GEO data were too nonstationary to yield a quality GW analysis, we searched for GWs in
coincidence with a total of six Parkes FRBs.  Analysis parameters were kept as similar as feasible to
the Green Bank drift-scan analysis described previously.  However, the upper end of the frequency range
was lowered to 1764 Hz due to the range over which GEO data are calibrated \cite{geogrb}
and the higher frequency
damped sinusoids were left out of the set of GW morphologies simulated.  Since the triggers are 
nominally at cosmological distances and we are unlikely to be sensitive to damped sinusoid-type signals
from neutron stars outside our own Galaxy, this limitation is not a major concern.

There was no evidence of a gravitational-wave signal for
any of these FRBs (the most significant single trial p-value was 0.07), although it should be noted that 
the smaller GEO interferometer is less sensitive
than the larger interferometers.  As such we treat this analysis primarily as a check for loud candidates 
and do not quote sensitivity upper limits for this search.  Instead a list of currently published FRBs without evidence for 
corresponding GW emission is given in Table \ref{FRBs}.
Since these are not consistent in terms of DM or other characteristics with the RRAT-like candidates
identified by Green Bank, we do not include these in the binomial test or other distributional studies
presented in Figs. 1 and 2.

\begin{center}
  \begin{table*}
  \caption{\label{GBT}Analyzed Green Bank Telescope single-pulse candidates}
  \begin{tabular}{l l l l l l |l l l}
  \hline
  Trigger & MJD & RA & Dec & DM & Interferometer & \hspace{1 pc}90\% C.L.  upper& limit  (h$_{\rm{rss}}$&$\times$10$^{-22}$ Hz$^{-\frac{1}{2}}$)\\
  Name & (geocentric) &  &      & (pc cm$^\mathrm{-3}$) & Network &150 Hz sinusoid & NS-NS  & 1750 Hz sinusoid\\
  \hline
  RRAT1704-0440 & 54240.38329 & 17$^\mathrm{h}$04$^\mathrm{m}$54.4$^\mathrm{s}$ & -4$^{\circ}$40$^\mathrm{'}$37.\hspace{-0.2 pc}$^{''}$ \hspace{-0.45 pc}9  & 43$\pm$1  & H1H2L1V1 & 3.51  & 4.08 & 18.6\\
  RRAT 1537+2350 & 54243.25237  & 15$^\mathrm{h}$37$^\mathrm{m}$47.4$^\mathrm{s}$  & 23$^{\circ}$50$^{'}$51.\hspace{-0.2 pc}$^{''}$ \hspace{-0.45 pc} \negmedspace 1  & 15$\pm$1 & H1H2L1V1 & 2.33 & 2.91 & 10.5\\
  RRAT 1636+0131 & 54317.05160 & 16$^\mathrm{h}$36$^\mathrm{m}$19.4$^\mathrm{s}$ & 01$^{\circ}$31$^{'}$07.\hspace{-0.2 pc}$^{''}$ \hspace{-0.45 pc}1  & 30$\pm$2 & H1H2L1V1 & 3.03  & 3.60 & 31.7\\
  RRAT 1651+0130 & 54317.06177 & 16$^\mathrm{h}$51$^\mathrm{m}$42.0$^\mathrm{s}$ & 01$^{\circ}$30$^{'}$57.\hspace{-0.2 pc}$^{''}$ \hspace{-0.45 pc}4 & 29$\pm$1 & H1H2L1V1 & 3.21 & 3.91 & 27.8\\
  RRAT 0206-0443 & 54239.76133 & 02$^\mathrm{h}$06$^\mathrm{m}$11.0$^\mathrm{s}$ & -04$^{\circ}$43$^{'}$33.\hspace{-0.2 pc}$^{''}$ \hspace{-0.45 pc}4 & 15$\pm$1 & H1H2L1V1 & 4.17 & 5.29 & 23.2\\
  RRAT 0801-0746 & 54289.78625 & 08$^\mathrm{h}$01$^\mathrm{m}$37.5$^\mathrm{s}$ & -07$^{\circ}$47$^{'}$03.\hspace{-0.2 pc}$^{''}$ \hspace{-0.45 pc}2 & 38$\pm$1 & H1H2L1V1 & 3.85 & 4.24 & 193.\\
  RRAT 0808-0746 & 54289.79101 & 08$^\mathrm{h}$08$^\mathrm{m}$43.3$^\mathrm{s}$ & -07$^{\circ}$46$^{'}$58.\hspace{-0.2 pc}$^{''}$ \hspace{-0.45 pc}9 & 36$\pm$1 & H1H2L1V1 & 3.55 & 4.32 & 72.5\\
  RRAT 0858-0746 & 54289.82543 & 08$^\mathrm{h}$58$^\mathrm{m}$23.9$^\mathrm{s}$  & -07$^{\circ}$46$^{'}$31.\hspace{-0.2 pc}$^{''}$ \hspace{-0.45 pc}4 & 40$\pm$3 & H1H2L1V1 & 4.35 & 4.32 & 14.4\\
  RRAT 0719-1907 & 54296.74455 & 07$^\mathrm{h}$19$^\mathrm{m}$38.3$^\mathrm{s}$ & -19$^{\circ}$07$^{'}$30.\hspace{-0.2 pc}$^{''}$ \hspace{-0.45 pc}2 & 27$\pm$1 & H1H2L1V1 & 3.96 & 4.66 & 21.8\\
  RRAT 2324-0507 & 54240.64681 & 23$^\mathrm{h}$24$^\mathrm{m}$22.2$^\mathrm{s}$ & -05$^{\circ}$07$^{'}$36.\hspace{-0.2 pc}$^{''}$ \hspace{-0.45 pc}0 & 15$\pm$1 & H1H2V1 & 6.09 & 5.95 & 28.6\\
  RRAT 2006+2021 & 54272.36809 & 20$^\mathrm{h}$06$^\mathrm{m}$54.8$^\mathrm{s}$ & 20$^{\circ}$21$^{'}$05.\hspace{-0.2 pc}$^{''}$ \hspace{-0.45 pc}8 & 66$\pm$1 & H1H2V1 & 6.01 & 6.77 & 17.4\\
  RRAT 1610-0128 & 54268.26252 & 16$^\mathrm{h}$10$^\mathrm{m}$58.4$^\mathrm{s}$ & -01$^{\circ}$28$^{'}$04.\hspace{-0.2 pc}$^{''}$ \hspace{-0.45 pc}0 & 27$\pm$1 & H1H2V1 & 5.90 & 6.71 & 21.7\\
  RRAT 0706-1058 & 54315.66206 & 07$^\mathrm{h}$06$^\mathrm{m}$43.2$^\mathrm{s}$ & -10$^{\circ}$58$^{'}$20.\hspace{-0.2 pc}$^{''}$ \hspace{-0.45 pc}3 & 19$\pm$1 & H1H2V1 & 5.21 & 5.79 & 25.6\\
  RRAT 0606-0129 & 54267.84394 & 06$^\mathrm{h}$06$^\mathrm{m}$37.5$^\mathrm{s}$ & -01$^{\circ}$29$^{'}$23.\hspace{-0.2 pc}$^{''}$ \hspace{-0.45 pc}9 & 17$\pm$3 & H1H2V1 & 5.86 & 5.89 & 25.2\\
  RRAT 0645-0128 & 54267.87160 & 06$^\mathrm{h}$45$^\mathrm{m}$39.5$^\mathrm{s}$ & -01$^{\circ}$28$^{'}$58.\hspace{-0.2 pc}$^{''}$ \hspace{-0.45 pc}1 & 15$\pm$2 & H1H2V1 & 5.14 & 5.65 & 24.2\\
  RRAT 1336-2034 & 54294.01362 & 13$^\mathrm{h}$36$^\mathrm{m}$28.2$^\mathrm{s}$ & -20$^{\circ}$34$^{'}$21.\hspace{-0.2 pc}$^{''}$ \hspace{-0.45 pc}8 & 19$\pm$1 & H1H2V1 & 5.63 & 5.98 & 25.1\\
  RRAT 0526-1908 & 54296.66537 & 05$^\mathrm{h}$26$^\mathrm{m}$04.9$^\mathrm{s}$ & -19$^{\circ}$08$^{'}$48.\hspace{-0.2 pc}$^{''}$ \hspace{-0.45 pc}8 & 26$\pm$2 & H1H2V1 & 5.64 & 5.90 & 25.0\\
  RRAT 1926+2021 & 54272.33975 & 19$^\mathrm{h}$26$^\mathrm{m}$41.9$^\mathrm{s}$ & 20$^{\circ}$21$^{'}$29.\hspace{-0.2 pc}$^{''}$ \hspace{-0.45 pc}1 & 21$\pm$1 & H2L1V1 & 7.39 & 115. & 17.5\\
  RRAT 1914-1129 & 54315.16846 & 19$^\mathrm{h}$14$^\mathrm{m}$46.3$^\mathrm{s}$ & -11$^{\circ}$29$^{'}$33.\hspace{-0.2 pc}$^{''}$ \hspace{-0.45 pc}5 & 91$\pm$1 & H1H2L1 & 3.01 & 3.32 & 15.4\\
  RRAT 1132+2455 & 54228.11976 & 11$^\mathrm{h}$32$^\mathrm{m}$09.2$^\mathrm{s}$ & 24$^{\circ}$55$^{'}$58.\hspace{-0.2 pc}$^{''}$ \hspace{-0.45 pc}7 & 24$\pm$1 & H1H2L1 & 2.65 & 2.90 & 12.8\\
  RRAT 1059-0102 & 54274.03007 & 10$^\mathrm{h}$59$^\mathrm{m}$31.9$^\mathrm{s}$ & -01$^{\circ}$02$^{'}$01.\hspace{-0.2 pc}$^{''}$ \hspace{-0.45 pc}6 & 18$\pm$1 & H1L1V1 & 4.35 & 5.67 & 70.5\\
  RRAT 1944-1017 & 54281.29872 & 19$^\mathrm{h}$44$^\mathrm{m}$08.6$^\mathrm{s}$ & -10$^{\circ}$17$^{'}$07.\hspace{-0.2 pc}$^{''}$ \hspace{-0.45 pc}2 & 31$\pm$1 & H1H2L1 & 2.84 & 3.33 & 15.6\\
  RRAT 0807-1057 & 54315.70287 & 08$^\mathrm{h}$07$^\mathrm{m}$02.7$^\mathrm{s}$ & -10$^{\circ}$57$^{'}$41.\hspace{-0.2 pc}$^{''}$ \hspace{-0.45 pc}4 & 18$\pm$1 & H1H2 & 5.66 & 5.39 & 22.8\\
  RRAT 0614-0328 & 54224.97146 & 6$^\mathrm{h}$14$^\mathrm{m}$37.8$^\mathrm{s}$ & -3$^{\circ}$28$^{'}$44.\hspace{-0.2 pc}$^{''}$ \hspace{-0.45 pc}9 & 18$\pm$1 & H1L1 & 2.50 & 3.06 & 1062\\
  RRAT 0544-0309 & 54226.94475 & 5$^\mathrm{h}$44$^\mathrm{m}$46.3$^\mathrm{s}$ & -3$^{\circ}$9$^{'}$44.\hspace{-0.2 pc}$^{''}$ \hspace{-0.45 pc}8 & 67$\pm$1 & H1L1 & 2.51 & 3.05 & 165.0\\
  GBNCC 04413 & 55163.27053 & 2$^\mathrm{h}$3$^\mathrm{m}$27.6$^\mathrm{s}$ & 70$^{\circ}$22$^{'}$43.\hspace{-0.2 pc}$^{''}$ \hspace{-0.45 pc}7 & 21$\pm$1 & H1V1 & 5.78 & 5.55 & 328.0\\
  GBNCC 04743 & 55169.16190 & 0$^\mathrm{h}$53$^\mathrm{m}$24.0$^\mathrm{s}$ & 69$^{\circ}$38$^{'}$45.\hspace{-0.2 pc}$^{''}$ \hspace{-0.45 pc}2 & 90$\pm$1 & H1V1 & 8.43 & 9.03 & 138.0\\

  \end{tabular}
  \end{table*}
\end{center}  

\begin{center}
  \begin{table*}
  \caption{\label{FRBs}Analyzed FRB candidates from the Parkes telescope. No
evidence of gravitational-wave emission was observed in coincidence with these FRBs.}
  \begin{tabular}{l l l l l l}
  \hline
  Trigger & MJD & RA & Dec & DM & Interferometer \\
  Name & (geocentric) &  &      & (pc cm$^\mathrm{-3}$) & Network\\
  \hline
  FRB 110626 & 55738.89810 & 21$^\mathrm{h}$03$^\mathrm{m}$43$^\mathrm{s}$ & -44$^{\circ}$44$^{'}$19$^{''}$  & 723.0  & G1V1\\
  FRB 110703 & 55745.79142 & 23$^\mathrm{h}$30$^\mathrm{m}$51$^\mathrm{s}$  & -02$^{\circ}$52$^{'}$24$^{''}$  & 1103.6 & G1V1 \\
  FRB 110220 & 55612.08041 & 22$^\mathrm{h}$34$^\mathrm{m}$38.2$^\mathrm{s}$ & -12$^{\circ}$33$^{'}$44$^{''}$  & 944.8  & G1 \\
  FRB 120127 & 55953.34122 & 23$^\mathrm{h}$15$^\mathrm{m}$6.3$^\mathrm{s}$ & -18$^{\circ}$25$^{'}$37$^{''}$  & 555.2 & G1 \\
  FRB 130628 & 56471.16527 & 9$^\mathrm{h}$03$^\mathrm{m}$2.5$^\mathrm{s}$ & 3$^{\circ}$26$^{'}$16$^{''}$  & 469.7  & G1 \\
  FRB 131104 & 56600.75279 & 6$^\mathrm{h}$44$^\mathrm{m}$10.4$^\mathrm{s}$ & -51$^{\circ}$16$^{'}$40$^{''}$  & 779.3 & G1 \\
  \end{tabular}
  \end{table*}
\end{center}

\section{Conclusions and Future Prospects}

The 
searches described in this current paper should be viewed largely as prototypes for
future searches with instruments that will eventually be an order of
magnitude more sensitive than the best sensitivities presented here.
Since much is currently unknown about FRBs and related phenomena, 
identification of a GW in close coincidence with a radio burst could provide
insight into both the distance and, depending on the GW morphology, 
astrophysical origin of the radio transient.

In addition to more sensitive searches in coincidence with fast radio bursts, 
efforts are also underway within the LIGO
and Virgo collaborations to analyze radio transients of longer durations
resulting from instruments operating at lower frequency than the Green Bank or
Parkes telescopes.  Since these transients have properties very different
than the ones described here and are not generally expected to come from the 
same sources, substantially different analysis methods are
required to address searches for GWs in coincidence with these signals \cite{Yancey}.

In the case of fast radio bursts, it is worth noting that arguments based on the Parkes field of view and observation
time suggest that if FRBs are in fact of astrophysical origin, the vast majority of FRBs are currently
missed by radio telescopes \cite{parkesfrbs,parkesmidgalactic}.  Accounting for possible anisotropies
in the distribution of FRBs and including both mid-galactic and high-latitude survey data, the all-sky rate for FRBs
is estimated to be between 1100 to 9600 FRBs per day above a threshold fluence of 4.0 Jy ms \cite{parkeshilat}.  While
externally triggered searches can look for signals with amplitudes of as much as $\sim$3 lower than all-sky searches \cite{was},
if such a population of FRBs generated detectable gravitational-wave signals, statistical arguments suggest they would 
be likely to show up in all-sky transient searches (e.g. \cite{allsky}) as well. However, these detections would not
be clearly associated with a FRB and thus lack the ability of a multimessenger search in constraining the possible
source of FRBs.
In the coming years, statistical information on FRBs is likely
to be dramatically improved, especially as wider field radio instruments come online \cite{lorimerdist,nexgenradio,mwafrb}.  

\begin{acknowledgments}

The authors gratefully acknowledge the support of the United States
National Science Foundation (NSF) for the construction and operation of the
LIGO Laboratory and Advanced LIGO as well as the Science and Technology Facilities Council (STFC) of the
United Kingdom, the Max-Planck-Society (MPS), and the State of
Niedersachsen/Germany for support of the construction of Advanced LIGO 
and construction and operation of the GEO600 detector. 
Additional support for Advanced LIGO was provided by the Australian Research Council.
The authors gratefully acknowledge the Italian Istituto Nazionale di Fisica Nucleare (INFN),  
the French Centre National de la Recherche Scientifique (CNRS) and
the Foundation for Fundamental Research on Matter supported by the Netherlands Organisation for Scientific Research, 
for the construction and operation of the Virgo detector
and the creation and support  of the EGO consortium. 
The authors also gratefully acknowledge research support from these agencies as well as by 
the Council of Scientific and Industrial Research of India, 
Department of Science and Technology, India,
Science \& Engineering Research Board (SERB), India,
Ministry of Human Resource Development, India,
the Spanish Ministerio de Econom\'ia y Competitividad,
the Conselleria d'Economia i Competitivitat and Conselleria d'Educaci\'o, Cultura i Universitats of the Govern de les Illes Balears,
the National Science Centre of Poland,
the European Commission,
the Royal Society, 
the Scottish Funding Council, 
the Scottish Universities Physics Alliance, 
the Hungarian Scientific Research Fund (OTKA),
the Lyon Institute of Origins (LIO),
the National Research Foundation of Korea,
Industry Canada and the Province of Ontario through the Ministry of Economic Development and Innovation, 
the Natural Science and Engineering Research Council Canada,
Canadian Institute for Advanced Research,
the Brazilian Ministry of Science, Technology, and Innovation,
Russian Foundation for Basic Research,
the Leverhulme Trust, 
the Research Corporation, 
Ministry of Science and Technology (MOST), Taiwan
and
the Kavli Foundation.
Some authors were supported by the European Research Council, including Grant No. 617199 to J.v.L.
The authors gratefully acknowledge the support of the NSF, STFC, MPS, INFN, CNRS and the
State of Niedersachsen/Germany for provision of computational resources.

\end{acknowledgments}

\bibliography{references}{}

\begin{thebibliography}{72}%
\makeatletter
\providecommand \@ifxundefined [1]{%
 \@ifx{#1\undefined}
}%
\providecommand \@ifnum [1]{%
 \ifnum #1\expandafter \@firstoftwo
 \else \expandafter \@secondoftwo
 \fi
}%
\providecommand \@ifx [1]{%
 \ifx #1\expandafter \@firstoftwo
 \else \expandafter \@secondoftwo
 \fi
}%
\providecommand \natexlab [1]{#1}%
\providecommand \enquote  [1]{``#1''}%
\providecommand \bibnamefont  [1]{#1}%
\providecommand \bibfnamefont [1]{#1}%
\providecommand \citenamefont [1]{#1}%
\providecommand \href@noop [0]{\@secondoftwo}%
\providecommand \href [0]{\begingroup \@sanitize@url \@href}%
\providecommand \@href[1]{\@@startlink{#1}\@@href}%
\providecommand \@@href[1]{\endgroup#1\@@endlink}%
\providecommand \@sanitize@url [0]{\catcode `\\12\catcode `\$12\catcode
  `\&12\catcode `\#12\catcode `\^12\catcode `\_12\catcode `\%12\relax}%
\providecommand \@@startlink[1]{}%
\providecommand \@@endlink[0]{}%
\providecommand \url  [0]{\begingroup\@sanitize@url \@url }%
\providecommand \@url [1]{\endgroup\@href {#1}{\urlprefix }}%
\providecommand \urlprefix  [0]{URL }%
\providecommand \Eprint [0]{\href }%
\providecommand \doibase [0]{http://dx.doi.org/}%
\providecommand \selectlanguage [0]{\@gobble}%
\providecommand \bibinfo  [0]{\@secondoftwo}%
\providecommand \bibfield  [0]{\@secondoftwo}%
\providecommand \translation [1]{[#1]}%
\providecommand \BibitemOpen [0]{}%
\providecommand \bibitemStop [0]{}%
\providecommand \bibitemNoStop [0]{.\EOS\space}%
\providecommand \EOS [0]{\spacefactor3000\relax}%
\providecommand \BibitemShut  [1]{\csname bibitem#1\endcsname}%
\let\auto@bib@innerbib\@empty
\bibitem [{aVi()}]{aVirgo}%
  \BibitemOpen
  \href@noop {} {\bibinfo  {journal}
  {https://wwwcascina.virgo.infn.it/advirgo/docs.html}\ }\BibitemShut {NoStop}%
\bibitem [{\citenamefont {Aasi}\ \emph {et~al.}(2013)\citenamefont {Aasi} \emph
  {et~al.}}]{pulsar2}%
  \BibitemOpen
\bibfield  {journal} {  }\bibfield  {author} {\bibinfo {author} {\bibnamefont
  {Aasi}, \bibfnamefont {J.}} \emph {et~al.},\ }\href@noop {} {\bibfield
  {journal} {\bibinfo  {journal} {Phys. Rev. D}\ }\textbf {\bibinfo {volume}
  {88}},\ \bibinfo {pages} {102002} (\bibinfo {year} {2013})}\BibitemShut
  {NoStop}%
\bibitem [{\citenamefont {Aasi}\ \emph
  {et~al.}(2014{\natexlab{a}})\citenamefont {Aasi} \emph {et~al.}}]{pulsar1}%
  \BibitemOpen
  \bibfield  {author} {\bibinfo {author} {\bibnamefont {Aasi}, \bibfnamefont
  {J.}} \emph {et~al.},\ }\href@noop {} {\bibfield  {journal} {\bibinfo
  {journal} {Astrophys. J.}\ }\textbf {\bibinfo {volume} {785}},\ \bibinfo
  {pages} {119} (\bibinfo {year} {2014}{\natexlab{a}})}\BibitemShut {NoStop}%
\bibitem [{\citenamefont {Aasi}\ \emph
  {et~al.}(2014{\natexlab{b}})\citenamefont {Aasi} \emph {et~al.}}]{geogrb}%
  \BibitemOpen
  \bibfield  {author} {\bibinfo {author} {\bibnamefont {Aasi}, \bibfnamefont
  {J.}} \emph {et~al.},\ }\href@noop {} {\bibfield  {journal} {\bibinfo
  {journal} {Phys. Rev. D}\ }\textbf {\bibinfo {volume} {89}},\ \bibinfo
  {pages} {122004} (\bibinfo {year} {2014}{\natexlab{b}})}\BibitemShut
  {NoStop}%
\bibitem [{\citenamefont {Aasi}\ \emph
  {et~al.}(2014{\natexlab{c}})\citenamefont {Aasi} \emph {et~al.}}]{cssearch}%
  \BibitemOpen
  \bibfield  {author} {\bibinfo {author} {\bibnamefont {Aasi}, \bibfnamefont
  {J.}} \emph {et~al.},\ }\href@noop {} {\bibfield  {journal} {\bibinfo
  {journal} {Phys. Rev. Lett.}\ }\textbf {\bibinfo {volume} {112}},\ \bibinfo
  {pages} {131101} (\bibinfo {year} {2014}{\natexlab{c}})}\BibitemShut
  {NoStop}%
\bibitem [{\citenamefont {Aasi}\ \emph
  {et~al.}(2014{\natexlab{d}})\citenamefont {Aasi} \emph
  {et~al.}}]{ipngrbsearch}%
  \BibitemOpen
  \bibfield  {author} {\bibinfo {author} {\bibnamefont {Aasi}, \bibfnamefont
  {J.}} \emph {et~al.},\ }\href@noop {} {\bibfield  {journal} {\bibinfo
  {journal} {Phys. Rev. Lett.}\ }\textbf {\bibinfo {volume} {113}},\ \bibinfo
  {pages} {011102} (\bibinfo {year} {2014}{\natexlab{d}})}\BibitemShut
  {NoStop}%
\bibitem [{\citenamefont {Aasi}\ \emph {et~al.}(2015)\citenamefont {Aasi} \emph
  {et~al.}}]{aLIGO}%
  \BibitemOpen
  \bibfield  {author} {\bibinfo {author} {\bibnamefont {Aasi}, \bibfnamefont
  {J.}} \emph {et~al.},\ }\href@noop {} {\bibfield  {journal} {\bibinfo
  {journal} {Class. Quant. Grav.}\ }\textbf {\bibinfo {volume} {32}},\ \bibinfo
  {pages} {074001} (\bibinfo {year} {2015})}\BibitemShut {NoStop}%
\bibitem [{\citenamefont {Abadie}\ \emph
  {et~al.}(2010{\natexlab{a}})\citenamefont {Abadie} \emph
  {et~al.}}]{cbcrates}%
  \BibitemOpen
  \bibfield  {author} {\bibinfo {author} {\bibnamefont {Abadie}, \bibfnamefont
  {J.}} \emph {et~al.},\ }\href@noop {} {\bibfield  {journal} {\bibinfo
  {journal} {Class. Quant. Grav.}\ }\textbf {\bibinfo {volume} {27}},\ \bibinfo
  {pages} {173001} (\bibinfo {year} {2010}{\natexlab{a}})}\BibitemShut
  {NoStop}%
\bibitem [{\citenamefont {Abadie}\ \emph
  {et~al.}(2010{\natexlab{b}})\citenamefont {Abadie} \emph
  {et~al.}}]{calibrationS5}%
  \BibitemOpen
  \bibfield  {author} {\bibinfo {author} {\bibnamefont {Abadie}, \bibfnamefont
  {J.}} \emph {et~al.},\ }\href@noop {} {\bibfield  {journal} {\bibinfo
  {journal} {Nucl. Inst. Meth. Phys. Res}\ }\textbf {\bibinfo {volume}
  {A624}},\ \bibinfo {pages} {223} (\bibinfo {year}
  {2010}{\natexlab{b}})}\BibitemShut {NoStop}%
\bibitem [{\citenamefont {Abadie}\ \emph {et~al.}(2011)\citenamefont {Abadie}
  \emph {et~al.}}]{clark}%
  \BibitemOpen
  \bibfield  {author} {\bibinfo {author} {\bibnamefont {Abadie}, \bibfnamefont
  {J.}} \emph {et~al.},\ }\href@noop {} {\bibfield  {journal} {\bibinfo
  {journal} {Phys. Rev. D}\ }\textbf {\bibinfo {volume} {83}},\ \bibinfo
  {pages} {042001} (\bibinfo {year} {2011})}\BibitemShut {NoStop}%
\bibitem [{\citenamefont {Abadie}\ \emph
  {et~al.}(2012{\natexlab{a}})\citenamefont {Abadie} \emph
  {et~al.}}]{grbsearch}%
  \BibitemOpen
  \bibfield  {author} {\bibinfo {author} {\bibnamefont {Abadie}, \bibfnamefont
  {J.}} \emph {et~al.},\ }\href@noop {} {\bibfield  {journal} {\bibinfo
  {journal} {Astrophys. J.}\ }\textbf {\bibinfo {volume} {760}},\ \bibinfo
  {pages} {12} (\bibinfo {year} {2012}{\natexlab{a}})}\BibitemShut {NoStop}%
\bibitem [{\citenamefont {Abadie}\ \emph
  {et~al.}(2012{\natexlab{b}})\citenamefont {Abadie} \emph {et~al.}}]{allsky}%
  \BibitemOpen
  \bibfield  {author} {\bibinfo {author} {\bibnamefont {Abadie}, \bibfnamefont
  {J.}} \emph {et~al.},\ }\href@noop {} {\bibfield  {journal} {\bibinfo
  {journal} {Phys. Rev. D}\ }\textbf {\bibinfo {volume} {85}},\ \bibinfo
  {pages} {122007} (\bibinfo {year} {2012}{\natexlab{b}})}\BibitemShut
  {NoStop}%
\bibitem [{\citenamefont {Abbott}\ \emph
  {et~al.}(2008{\natexlab{a}})\citenamefont {Abbott} \emph {et~al.}}]{SGR}%
  \BibitemOpen
  \bibfield  {author} {\bibinfo {author} {\bibnamefont {Abbott}, \bibfnamefont
  {B.~P.}} \emph {et~al.},\ }\href@noop {} {\bibfield  {journal} {\bibinfo
  {journal} {Phys. Rev. Lett.}\ }\textbf {\bibinfo {volume} {101}},\ \bibinfo
  {pages} {211102} (\bibinfo {year} {2008}{\natexlab{a}})}\BibitemShut
  {NoStop}%
\bibitem [{\citenamefont {Abbott}\ \emph
  {et~al.}(2008{\natexlab{b}})\citenamefont {Abbott} \emph {et~al.}}]{GRB1}%
  \BibitemOpen
  \bibfield  {author} {\bibinfo {author} {\bibnamefont {Abbott}, \bibfnamefont
  {B.~P.}} \emph {et~al.},\ }\href@noop {} {\bibfield  {journal} {\bibinfo
  {journal} {Phys. Rev. D}\ }\textbf {\bibinfo {volume} {77}},\ \bibinfo
  {pages} {062004} (\bibinfo {year} {2008}{\natexlab{b}})}\BibitemShut
  {NoStop}%
\bibitem [{\citenamefont {Abbott}\ \emph {et~al.}(2009)\citenamefont {Abbott}
  \emph {et~al.}}]{LIGO}%
  \BibitemOpen
  \bibfield  {author} {\bibinfo {author} {\bibnamefont {Abbott}, \bibfnamefont
  {B.~P.}} \emph {et~al.},\ }\href@noop {} {\bibfield  {journal} {\bibinfo
  {journal} {Reports on Progress in Physics}\ }\textbf {\bibinfo {volume}
  {72}},\ \bibinfo {pages} {076901} (\bibinfo {year} {2009})}\BibitemShut
  {NoStop}%
\bibitem [{\citenamefont {Abbott}\ \emph {et~al.}(2016)\citenamefont {Abbott}
  \emph {et~al.}}]{discovery}%
  \BibitemOpen
  \bibfield  {author} {\bibinfo {author} {\bibnamefont {Abbott}, \bibfnamefont
  {B.~P.}} \emph {et~al.},\ }\href@noop {} {\bibfield  {journal} {\bibinfo
  {journal} {Phys. Rev. Lett.}\ }\textbf {\bibinfo {volume} {116}},\ \bibinfo
  {pages} {061102} (\bibinfo {year} {2016})}\BibitemShut {NoStop}%
\bibitem [{\citenamefont {Accadia}\ \emph {et~al.}(2012)\citenamefont {Accadia}
  \emph {et~al.}}]{Virgo}%
  \BibitemOpen
  \bibfield  {author} {\bibinfo {author} {\bibnamefont {Accadia}, \bibfnamefont
  {T.}} \emph {et~al.},\ }\href@noop {} {\bibfield  {journal} {\bibinfo
  {journal} {Journal of Instrumentation}\ }\textbf {\bibinfo {volume} {7}},\
  \bibinfo {pages} {P03012} (\bibinfo {year} {2012})}\BibitemShut {NoStop}%
\bibitem [{\citenamefont {Bannister}\ and\ \citenamefont
  {Madsen}(2014)}]{FRBgalactic}%
  \BibitemOpen
  \bibfield  {author} {\bibinfo {author} {\bibnamefont {Bannister},
  \bibfnamefont {K.~W.}}\ and\ \bibinfo {author} {\bibnamefont {Madsen},
  \bibfnamefont {G.~J.}},\ }\href@noop {} {\bibfield  {journal} {\bibinfo
  {journal} {Mon. Not. R. Astron. Soc.}\ }\textbf {\bibinfo {volume} {440}},\
  \bibinfo {pages} {353} (\bibinfo {year} {2014})}\BibitemShut {NoStop}%
\bibitem [{\citenamefont {Bannister}\ \emph {et~al.}(2012)\citenamefont
  {Bannister}, \citenamefont {Murphy}, \citenamefont {Gaensler},\ and\
  \citenamefont {Reynolds}}]{bannister}%
  \BibitemOpen
  \bibfield  {author} {\bibinfo {author} {\bibnamefont {Bannister},
  \bibfnamefont {K.~W.}}, \bibinfo {author} {\bibnamefont {Murphy},
  \bibfnamefont {T.}}, \bibinfo {author} {\bibnamefont {Gaensler},
  \bibfnamefont {B.~M.}}, \ and\ \bibinfo {author} {\bibnamefont {Reynolds},
  \bibfnamefont {J.~E.}},\ }\href@noop {} {\bibfield  {journal} {\bibinfo
  {journal} {Astrophys. J.}\ }\textbf {\bibinfo {volume} {757}},\ \bibinfo
  {pages} {38} (\bibinfo {year} {2012})}\BibitemShut {NoStop}%
\bibitem [{\citenamefont {Benhar}, \citenamefont {Ferrari},\ and\ \citenamefont
  {Gualtieri}(2004)}]{fmode}%
  \BibitemOpen
  \bibfield  {author} {\bibinfo {author} {\bibnamefont {Benhar}, \bibfnamefont
  {O.}}, \bibinfo {author} {\bibnamefont {Ferrari}, \bibfnamefont {V.}}, \ and\
  \bibinfo {author} {\bibnamefont {Gualtieri}, \bibfnamefont {L.}},\
  }\href@noop {} {\bibfield  {journal} {\bibinfo  {journal} {Phys. Rev. D}\
  }\textbf {\bibinfo {volume} {70}},\ \bibinfo {pages} {124015} (\bibinfo
  {year} {2004})}\BibitemShut {NoStop}%
\bibitem [{\citenamefont {Berezinsky}, \citenamefont {Hnatyk},\ and\
  \citenamefont {Vilenkin}(2004)}]{berezinsky}%
  \BibitemOpen
  \bibfield  {author} {\bibinfo {author} {\bibnamefont {Berezinsky},
  \bibfnamefont {V.}}, \bibinfo {author} {\bibnamefont {Hnatyk}, \bibfnamefont
  {B.}}, \ and\ \bibinfo {author} {\bibnamefont {Vilenkin}, \bibfnamefont
  {A.}},\ }\href@noop {} {\bibfield  {journal} {\bibinfo  {journal} {Baltic
  Astronomy}\ }\textbf {\bibinfo {volume} {13}},\ \bibinfo {pages} {289}
  (\bibinfo {year} {2004})}\BibitemShut {NoStop}%
\bibitem [{\citenamefont {Boyles}\ \emph {et~al.}(2013)\citenamefont {Boyles}
  \emph {et~al.}}]{gbtdriftscan}%
  \BibitemOpen
  \bibfield  {author} {\bibinfo {author} {\bibnamefont {Boyles}, \bibfnamefont
  {J.}} \emph {et~al.},\ }\href@noop {} {\bibfield  {journal} {\bibinfo
  {journal} {Astrophys. J.}\ }\textbf {\bibinfo {volume} {763}},\ \bibinfo
  {pages} {80} (\bibinfo {year} {2013})}\BibitemShut {NoStop}%
\bibitem [{\citenamefont {Cai}, \citenamefont {Sabancilar},\ and\ \citenamefont
  {Vachaspati}(2012)}]{cai}%
  \BibitemOpen
  \bibfield  {author} {\bibinfo {author} {\bibnamefont {Cai}, \bibfnamefont
  {Y.~F.}}, \bibinfo {author} {\bibnamefont {Sabancilar}, \bibfnamefont {E.}},
  \ and\ \bibinfo {author} {\bibnamefont {Vachaspati}, \bibfnamefont {T.}},\
  }\href@noop {} {\bibfield  {journal} {\bibinfo  {journal} {Phys. Rev. D}\
  }\textbf {\bibinfo {volume} {85}},\ \bibinfo {pages} {023530} (\bibinfo
  {year} {2012})}\BibitemShut {NoStop}%
\bibitem [{\citenamefont {Damour}\ and\ \citenamefont
  {Vilenkin}(2000)}]{damour}%
  \BibitemOpen
  \bibfield  {author} {\bibinfo {author} {\bibnamefont {Damour}, \bibfnamefont
  {T.}}\ and\ \bibinfo {author} {\bibnamefont {Vilenkin}, \bibfnamefont {A.}},\
  }\href@noop {} {\bibfield  {journal} {\bibinfo  {journal} {Phys. Rev. Lett.}\
  }\textbf {\bibinfo {volume} {85}},\ \bibinfo {pages} {3761} (\bibinfo {year}
  {2000})}\BibitemShut {NoStop}%
\bibitem [{\citenamefont {Duncan}(1998)}]{duncan}%
  \BibitemOpen
  \bibfield  {author} {\bibinfo {author} {\bibnamefont {Duncan}, \bibfnamefont
  {R.~C.}},\ }\href@noop {} {\bibfield  {journal} {\bibinfo  {journal}
  {Astrophys. J. Lett.}\ }\textbf {\bibinfo {volume} {498}},\ \bibinfo {pages}
  {L45} (\bibinfo {year} {1998})}\BibitemShut {NoStop}%
\bibitem [{\citenamefont {Franco}, \citenamefont {Link},\ and\ \citenamefont
  {Epstein}(2000)}]{starquakes}%
  \BibitemOpen
  \bibfield  {author} {\bibinfo {author} {\bibnamefont {Franco}, \bibfnamefont
  {L.~M.}}, \bibinfo {author} {\bibnamefont {Link}, \bibfnamefont {B.}}, \ and\
  \bibinfo {author} {\bibnamefont {Epstein}, \bibfnamefont {R.~I.}},\
  }\href@noop {} {\bibfield  {journal} {\bibinfo  {journal} {Astrophys. J.}\
  }\textbf {\bibinfo {volume} {543}},\ \bibinfo {pages} {987} (\bibinfo {year}
  {2000})}\BibitemShut {NoStop}%
\bibitem [{\citenamefont {Gao}, \citenamefont {Li},\ and\ \citenamefont
  {Zhang}(2014)}]{frbgrb}%
  \BibitemOpen
  \bibfield  {author} {\bibinfo {author} {\bibnamefont {Gao}, \bibfnamefont
  {H.}}, \bibinfo {author} {\bibnamefont {Li}, \bibfnamefont {Z.}}, \ and\
  \bibinfo {author} {\bibnamefont {Zhang}, \bibfnamefont {B.}},\ }\href@noop {}
  {\bibfield  {journal} {\bibinfo  {journal} {Astrophys J.}\ }\textbf {\bibinfo
  {volume} {788}},\ \bibinfo {pages} {189} (\bibinfo {year}
  {2014})}\BibitemShut {NoStop}%
\bibitem [{\citenamefont {Grote}\ \emph {et~al.}(2010)\citenamefont {Grote}
  \emph {et~al.}}]{GEO}%
  \BibitemOpen
  \bibfield  {author} {\bibinfo {author} {\bibnamefont {Grote}, \bibfnamefont
  {H.}} \emph {et~al.},\ }\href@noop {} {\bibfield  {journal} {\bibinfo
  {journal} {Class. Quant. Grav.}\ }\textbf {\bibinfo {volume} {27}},\ \bibinfo
  {pages} {084003} (\bibinfo {year} {2010})}\BibitemShut {NoStop}%
\bibitem [{\citenamefont {Hansen}\ and\ \citenamefont
  {Lyutikov}(2001)}]{hansen}%
  \BibitemOpen
  \bibfield  {author} {\bibinfo {author} {\bibnamefont {Hansen}, \bibfnamefont
  {B.~M.~S.}}\ and\ \bibinfo {author} {\bibnamefont {Lyutikov}, \bibfnamefont
  {M.}},\ }\href@noop {} {\bibfield  {journal} {\bibinfo  {journal} {Mon. Not.
  R. Astron. Soc.}\ }\textbf {\bibinfo {volume} {322}},\ \bibinfo {pages} {695}
  (\bibinfo {year} {2001})}\BibitemShut {NoStop}%
\bibitem [{\citenamefont {Hassall}, \citenamefont {Keane},\ and\ \citenamefont
  {Fender}(2013)}]{nexgenradio}%
  \BibitemOpen
  \bibfield  {author} {\bibinfo {author} {\bibnamefont {Hassall}, \bibfnamefont
  {T.~E.}}, \bibinfo {author} {\bibnamefont {Keane}, \bibfnamefont {E.~F.}}, \
  and\ \bibinfo {author} {\bibnamefont {Fender}, \bibfnamefont {R.~P.}},\
  }\href@noop {} {\bibfield  {journal} {\bibinfo  {journal} {Mon. Not. R.
  Astron. Soc.}\ }\textbf {\bibinfo {volume} {436}},\ \bibinfo {pages} {371}
  (\bibinfo {year} {2013})}\BibitemShut {NoStop}%
\bibitem [{\citenamefont {Iyer}\ \emph {et~al.}(2011)\citenamefont {Iyer} \emph
  {et~al.}}]{LIGOIndia}%
  \BibitemOpen
  \bibfield  {author} {\bibinfo {author} {\bibnamefont {Iyer}, \bibfnamefont
  {B.}} \emph {et~al.},\ }\href@noop {} {\bibfield  {journal} {\bibinfo
  {journal} {LIGO-India Tech. rep.
  https://dcc.ligo.org/cgi-bin/DocDB/ShowDocument?docid=75988}\ } (\bibinfo
  {year} {2011})}\BibitemShut {NoStop}%
\bibitem [{\citenamefont {Keane}\ \emph {et~al.}(2011)\citenamefont {Keane},
  \citenamefont {Kramer}, \citenamefont {Lyne}, \citenamefont {Stappers},\ and\
  \citenamefont {McLaughlin}}]{keane}%
  \BibitemOpen
  \bibfield  {author} {\bibinfo {author} {\bibnamefont {Keane}, \bibfnamefont
  {E.~F.}}, \bibinfo {author} {\bibnamefont {Kramer}, \bibfnamefont {M.}},
  \bibinfo {author} {\bibnamefont {Lyne}, \bibfnamefont {A.~G.}}, \bibinfo
  {author} {\bibnamefont {Stappers}, \bibfnamefont {B.~W.}}, \ and\ \bibinfo
  {author} {\bibnamefont {McLaughlin}, \bibfnamefont {M.~A.}},\ }\href@noop {}
  {\bibfield  {journal} {\bibinfo  {journal} {Mon. Not. R. Astron. Soc.}\
  }\textbf {\bibinfo {volume} {415}},\ \bibinfo {pages} {3065} (\bibinfo {year}
  {2011})}\BibitemShut {NoStop}%
\bibitem [{\citenamefont {Keane}\ \emph {et~al.}(2012)\citenamefont {Keane},
  \citenamefont {Stappers}, \citenamefont {Kramer},\ and\ \citenamefont
  {Lyne}}]{keaneburst}%
  \BibitemOpen
  \bibfield  {author} {\bibinfo {author} {\bibnamefont {Keane}, \bibfnamefont
  {E.~F.}}, \bibinfo {author} {\bibnamefont {Stappers}, \bibfnamefont {B.~W.}},
  \bibinfo {author} {\bibnamefont {Kramer}, \bibfnamefont {M.}}, \ and\
  \bibinfo {author} {\bibnamefont {Lyne}, \bibfnamefont {A.~G.}},\ }\href@noop
  {} {\bibfield  {journal} {\bibinfo  {journal} {Mon. Not. R. Astron. Soc.}\
  }\textbf {\bibinfo {volume} {425}},\ \bibinfo {pages} {L71} (\bibinfo {year}
  {2012})}\BibitemShut {NoStop}%
\bibitem [{\citenamefont {Keane}\ \emph {et~al.}(2016)\citenamefont {Keane}
  \emph {et~al.}}]{FRBhost}%
  \BibitemOpen
  \bibfield  {author} {\bibinfo {author} {\bibnamefont {Keane}, \bibfnamefont
  {E.~F.}} \emph {et~al.},\ }\href@noop {} {\bibfield  {journal} {\bibinfo
  {journal} {Nature}\ }\textbf {\bibinfo {volume} {530}},\ \bibinfo {pages}
  {453} (\bibinfo {year} {2016})}\BibitemShut {NoStop}%
\bibitem [{\citenamefont {Kokkotas}\ and\ \citenamefont
  {Schmidt}(1999)}]{kokkotas}%
  \BibitemOpen
  \bibfield  {author} {\bibinfo {author} {\bibnamefont {Kokkotas},
  \bibfnamefont {K.~D.}}\ and\ \bibinfo {author} {\bibnamefont {Schmidt},
  \bibfnamefont {B.~G.}},\ }\href@noop {} {\bibfield  {journal} {\bibinfo
  {journal} {Living Rev. Rel.}\ }\textbf {\bibinfo {volume} {2}},\ \bibinfo
  {pages} {2} (\bibinfo {year} {1999})}\BibitemShut {NoStop}%
\bibitem [{\citenamefont {Li}\ \emph {et~al.}(2014)\citenamefont {Li},
  \citenamefont {Zhou}, \citenamefont {He}, \citenamefont {Fan},\ and\
  \citenamefont {Wei}}]{FRBCR}%
  \BibitemOpen
  \bibfield  {author} {\bibinfo {author} {\bibnamefont {Li}, \bibfnamefont
  {X.}}, \bibinfo {author} {\bibnamefont {Zhou}, \bibfnamefont {B.}}, \bibinfo
  {author} {\bibnamefont {He}, \bibfnamefont {H.-N.}}, \bibinfo {author}
  {\bibnamefont {Fan}, \bibfnamefont {Y.-Z.}}, \ and\ \bibinfo {author}
  {\bibnamefont {Wei}, \bibfnamefont {D.~M.}},\ }\href@noop {} {\bibfield
  {journal} {\bibinfo  {journal} {Astrophys. J.}\ }\textbf {\bibinfo {volume}
  {797}},\ \bibinfo {pages} {33} (\bibinfo {year} {2014})}\BibitemShut
  {NoStop}%
\bibitem [{\citenamefont {Lipunov}\ and\ \citenamefont
  {Panchenko}(1996)}]{lipunov}%
  \BibitemOpen
  \bibfield  {author} {\bibinfo {author} {\bibnamefont {Lipunov}, \bibfnamefont
  {V.~M.}}\ and\ \bibinfo {author} {\bibnamefont {Panchenko}, \bibfnamefont
  {I.~E.}},\ }\href@noop {} {\bibfield  {journal} {\bibinfo  {journal}
  {Astonomy and Astrophysics}\ }\textbf {\bibinfo {volume} {312}},\ \bibinfo
  {pages} {937} (\bibinfo {year} {1996})}\BibitemShut {NoStop}%
\bibitem [{\citenamefont {Lipunov}\ and\ \citenamefont
  {Pruzhinskaya}(2014)}]{lipunov2}%
  \BibitemOpen
  \bibfield  {author} {\bibinfo {author} {\bibnamefont {Lipunov}, \bibfnamefont
  {V.~M.}}\ and\ \bibinfo {author} {\bibnamefont {Pruzhinskaya}, \bibfnamefont
  {M.~V.}},\ }\href@noop {} {\bibfield  {journal} {\bibinfo  {journal} {Mon.
  Not. R. Astron. Soc.}\ }\textbf {\bibinfo {volume} {440}},\ \bibinfo {pages}
  {1193} (\bibinfo {year} {2014})}\BibitemShut {NoStop}%
\bibitem [{\citenamefont {Lorimer}\ \emph {et~al.}(2007)\citenamefont
  {Lorimer}, \citenamefont {Bailes}, \citenamefont {McLaughlin}, \citenamefont
  {Narkevic},\ and\ \citenamefont {Crawford}}]{lorimerburst}%
  \BibitemOpen
  \bibfield  {author} {\bibinfo {author} {\bibnamefont {Lorimer}, \bibfnamefont
  {D.~R.}}, \bibinfo {author} {\bibnamefont {Bailes}, \bibfnamefont {M.}},
  \bibinfo {author} {\bibnamefont {McLaughlin}, \bibfnamefont {M.~A.}},
  \bibinfo {author} {\bibnamefont {Narkevic}, \bibfnamefont {D.~J.}}, \ and\
  \bibinfo {author} {\bibnamefont {Crawford}, \bibfnamefont {F.}},\ }\href@noop
  {} {\bibfield  {journal} {\bibinfo  {journal} {Science}\ }\textbf {\bibinfo
  {volume} {318}},\ \bibinfo {pages} {777} (\bibinfo {year}
  {2007})}\BibitemShut {NoStop}%
\bibitem [{\citenamefont {Lorimer}\ \emph {et~al.}(2013)\citenamefont
  {Lorimer}, \citenamefont {Karastergiou}, \citenamefont {McLaughlin},\ and\
  \citenamefont {Johnston}}]{lorimerdist}%
  \BibitemOpen
  \bibfield  {author} {\bibinfo {author} {\bibnamefont {Lorimer}, \bibfnamefont
  {D.~R.}}, \bibinfo {author} {\bibnamefont {Karastergiou}, \bibfnamefont
  {A.}}, \bibinfo {author} {\bibnamefont {McLaughlin}, \bibfnamefont {M.~A.}},
  \ and\ \bibinfo {author} {\bibnamefont {Johnston}, \bibfnamefont {S.}},\
  }\href@noop {} {\bibfield  {journal} {\bibinfo  {journal} {Mon. Not. R.
  Astron. Soc.}\ }\textbf {\bibinfo {volume} {436}},\ \bibinfo {pages} {L5}
  (\bibinfo {year} {2013})}\BibitemShut {NoStop}%
\bibitem [{\citenamefont {Lyne}\ \emph {et~al.}(2009)\citenamefont {Lyne} \emph
  {et~al.}}]{lyne}%
  \BibitemOpen
  \bibfield  {author} {\bibinfo {author} {\bibnamefont {Lyne}, \bibfnamefont
  {A.~G.}} \emph {et~al.},\ }\href@noop {} {\bibfield  {journal} {\bibinfo
  {journal} {Mon. Not. R. Astron. Soc.}\ }\textbf {\bibinfo {volume} {400}},\
  \bibinfo {pages} {1439} (\bibinfo {year} {2009})}\BibitemShut {NoStop}%
\bibitem [{\citenamefont {Masui}\ \emph {et~al.}(2015)\citenamefont {Masui}
  \emph {et~al.}}]{GBTFRB}%
  \BibitemOpen
  \bibfield  {author} {\bibinfo {author} {\bibnamefont {Masui}, \bibfnamefont
  {K.}} \emph {et~al.},\ }\href@noop {} {\bibfield  {journal} {\bibinfo
  {journal} {Nature}\ }\textbf {\bibinfo {volume} {528}},\ \bibinfo {pages}
  {523} (\bibinfo {year} {2015})}\BibitemShut {NoStop}%
\bibitem [{\citenamefont {McLaughlin}\ \emph {et~al.}(2006)\citenamefont
  {McLaughlin} \emph {et~al.}}]{RRATS}%
  \BibitemOpen
  \bibfield  {author} {\bibinfo {author} {\bibnamefont {McLaughlin},
  \bibfnamefont {M.~A.}} \emph {et~al.},\ }\href@noop {} {\bibfield  {journal}
  {\bibinfo  {journal} {Nature}\ }\textbf {\bibinfo {volume} {439}},\ \bibinfo
  {pages} {817} (\bibinfo {year} {2006})}\BibitemShut {NoStop}%
\bibitem [{\citenamefont {Middleditch}\ \emph {et~al.}(2006)\citenamefont
  {Middleditch}, \citenamefont {Marshall}, \citenamefont {Wang}, \citenamefont
  {Gotthelf},\ and\ \citenamefont {Zhang}}]{pulsarglitch}%
  \BibitemOpen
  \bibfield  {author} {\bibinfo {author} {\bibnamefont {Middleditch},
  \bibfnamefont {J.}}, \bibinfo {author} {\bibnamefont {Marshall},
  \bibfnamefont {F.~E.}}, \bibinfo {author} {\bibnamefont {Wang}, \bibfnamefont
  {Q.~D.}}, \bibinfo {author} {\bibnamefont {Gotthelf}, \bibfnamefont {E.~V.}},
  \ and\ \bibinfo {author} {\bibnamefont {Zhang}, \bibfnamefont {W.}},\
  }\href@noop {} {\bibfield  {journal} {\bibinfo  {journal} {Astrophys. J.}\
  }\textbf {\bibinfo {volume} {652}},\ \bibinfo {pages} {1531} (\bibinfo {year}
  {2006})}\BibitemShut {NoStop}%
\bibitem [{\citenamefont {Moortgat}\ and\ \citenamefont
  {Kuijpers}()}]{moortgat2}%
  \BibitemOpen
  \bibfield  {author} {\bibinfo {author} {\bibnamefont {Moortgat},
  \bibfnamefont {J.}}\ and\ \bibinfo {author} {\bibnamefont {Kuijpers},
  \bibfnamefont {J.}},\ }\href@noop {} {\bibinfo  {journal} {Proceedings of the
  XXII Texas Symposium on Relativistic Astrophysics (Stanford University,
  2004)}\ ,\ \bibinfo {pages} {PSN 1230}}\BibitemShut {NoStop}%
\bibitem [{\citenamefont {Moortgat}\ and\ \citenamefont
  {Kuijpers}(2004)}]{moortgat}%
  \BibitemOpen
\bibfield  {journal} {  }\bibfield  {author} {\bibinfo {author} {\bibnamefont
  {Moortgat}, \bibfnamefont {J.}}\ and\ \bibinfo {author} {\bibnamefont
  {Kuijpers}, \bibfnamefont {J.}},\ }\href@noop {} {\bibfield  {journal}
  {\bibinfo  {journal} {Phys. Rev. D}\ }\textbf {\bibinfo {volume} {70}},\
  \bibinfo {pages} {023001} (\bibinfo {year} {2004})}\BibitemShut {NoStop}%
\bibitem [{\citenamefont {Nakar}(2007)}]{nakar}%
  \BibitemOpen
  \bibfield  {author} {\bibinfo {author} {\bibnamefont {Nakar}, \bibfnamefont
  {E.}},\ }\href@noop {} {\bibfield  {journal} {\bibinfo  {journal} {Phys.
  Rep.}\ }\textbf {\bibinfo {volume} {442}},\ \bibinfo {pages} {166} (\bibinfo
  {year} {2007})}\BibitemShut {NoStop}%
\bibitem [{\citenamefont {Petroff}\ \emph {et~al.}()\citenamefont {Petroff}
  \emph {et~al.}}]{FRBCAT}%
  \BibitemOpen
  \bibfield  {author} {\bibinfo {author} {\bibnamefont {Petroff}, \bibfnamefont
  {E.}} \emph {et~al.},\ }\href@noop {} {\bibinfo  {journal}
  {arXiv:1601.03547}\ }\BibitemShut {NoStop}%
\bibitem [{\citenamefont {Petroff}\ \emph {et~al.}(2014)\citenamefont {Petroff}
  \emph {et~al.}}]{parkesmidgalactic}%
  \BibitemOpen
\bibfield  {journal} {  }\bibfield  {author} {\bibinfo {author} {\bibnamefont
  {Petroff}, \bibfnamefont {E.}} \emph {et~al.},\ }\href@noop {} {\bibfield
  {journal} {\bibinfo  {journal} {Astrophys. J.}\ }\textbf {\bibinfo {volume}
  {789}},\ \bibinfo {pages} {L26} (\bibinfo {year} {2014})}\BibitemShut
  {NoStop}%
\bibitem [{\citenamefont {Petroff}\ \emph
  {et~al.}(2015{\natexlab{a}})\citenamefont {Petroff} \emph
  {et~al.}}]{parkes140514}%
  \BibitemOpen
  \bibfield  {author} {\bibinfo {author} {\bibnamefont {Petroff}, \bibfnamefont
  {E.}} \emph {et~al.},\ }\href@noop {} {\bibfield  {journal} {\bibinfo
  {journal} {Mon. Not. R. Astron. Soc.}\ }\textbf {\bibinfo {volume} {447}},\
  \bibinfo {pages} {246} (\bibinfo {year} {2015}{\natexlab{a}})}\BibitemShut
  {NoStop}%
\bibitem [{\citenamefont {Petroff}\ \emph
  {et~al.}(2015{\natexlab{b}})\citenamefont {Petroff} \emph
  {et~al.}}]{microperyton}%
  \BibitemOpen
  \bibfield  {author} {\bibinfo {author} {\bibnamefont {Petroff}, \bibfnamefont
  {E.}} \emph {et~al.},\ }\href@noop {} {\bibfield  {journal} {\bibinfo
  {journal} {Mon. Not. R. Astron. Soc.}\ }\textbf {\bibinfo {volume} {451}},\
  \bibinfo {pages} {3933} (\bibinfo {year} {2015}{\natexlab{b}})}\BibitemShut
  {NoStop}%
\bibitem [{\citenamefont {Predoi}\ \emph {et~al.}(2010)\citenamefont {Predoi}
  \emph {et~al.}}]{predoi}%
  \BibitemOpen
  \bibfield  {author} {\bibinfo {author} {\bibnamefont {Predoi}, \bibfnamefont
  {V.}} \emph {et~al.},\ }\href@noop {} {\bibfield  {journal} {\bibinfo
  {journal} {Class. Quant. Grav.}\ }\textbf {\bibinfo {volume} {27}},\ \bibinfo
  {pages} {084018} (\bibinfo {year} {2010})}\BibitemShut {NoStop}%
\bibitem [{\citenamefont {Pshirkov}\ and\ \citenamefont
  {Postnov}(2010)}]{pshirkov}%
  \BibitemOpen
  \bibfield  {author} {\bibinfo {author} {\bibnamefont {Pshirkov},
  \bibfnamefont {M.}}\ and\ \bibinfo {author} {\bibnamefont {Postnov},
  \bibfnamefont {K.}},\ }\href@noop {} {\bibfield  {journal} {\bibinfo
  {journal} {Astrophysics and Space Science}\ }\textbf {\bibinfo {volume}
  {330}},\ \bibinfo {pages} {13} (\bibinfo {year} {2010})}\BibitemShut
  {NoStop}%
\bibitem [{\citenamefont {Rane}\ \emph {et~al.}(2016)\citenamefont {Rane} \emph
  {et~al.}}]{parkeshilat}%
  \BibitemOpen
  \bibfield  {author} {\bibinfo {author} {\bibnamefont {Rane}, \bibfnamefont
  {A.}} \emph {et~al.},\ }\href@noop {} {\bibfield  {journal} {\bibinfo
  {journal} {Mon. Not. R. Astron. Soc.}\ }\textbf {\bibinfo {volume} {455}},\
  \bibinfo {pages} {2207} (\bibinfo {year} {2016})}\BibitemShut {NoStop}%
\bibitem [{\citenamefont {Ravi}, \citenamefont {Shannon},\ and\ \citenamefont
  {Jameson}(2015)}]{parkes131104}%
  \BibitemOpen
  \bibfield  {author} {\bibinfo {author} {\bibnamefont {Ravi}, \bibfnamefont
  {V.}}, \bibinfo {author} {\bibnamefont {Shannon}, \bibfnamefont {R.}}, \ and\
  \bibinfo {author} {\bibnamefont {Jameson}, \bibfnamefont {A.}},\ }\href@noop
  {} {\bibfield  {journal} {\bibinfo  {journal} {Astrophys. J. Lett.}\ }\textbf
  {\bibinfo {volume} {799}},\ \bibinfo {pages} {L5} (\bibinfo {year}
  {2015})}\BibitemShut {NoStop}%
\bibitem [{\citenamefont {Rubio-Herrera}\ \emph {et~al.}(2013)\citenamefont
  {Rubio-Herrera}, \citenamefont {Stappers}, \citenamefont {Hessels},\ and\
  \citenamefont {Braun}}]{rubioherrera}%
  \BibitemOpen
  \bibfield  {author} {\bibinfo {author} {\bibnamefont {Rubio-Herrera},
  \bibfnamefont {E.}}, \bibinfo {author} {\bibnamefont {Stappers},
  \bibfnamefont {B.~W.}}, \bibinfo {author} {\bibnamefont {Hessels},
  \bibfnamefont {J.~W.~T.}}, \ and\ \bibinfo {author} {\bibnamefont {Braun},
  \bibfnamefont {R.}},\ }\href@noop {} {\bibfield  {journal} {\bibinfo
  {journal} {Mon. Not. R. Astron. Soc.}\ }\textbf {\bibinfo {volume} {428}},\
  \bibinfo {pages} {2857} (\bibinfo {year} {2013})}\BibitemShut {NoStop}%
\bibitem [{\citenamefont {Sagiv}\ and\ \citenamefont {Waxman}(2002)}]{sagiv}%
  \BibitemOpen
  \bibfield  {author} {\bibinfo {author} {\bibnamefont {Sagiv}, \bibfnamefont
  {A.}}\ and\ \bibinfo {author} {\bibnamefont {Waxman}, \bibfnamefont {E.}},\
  }\href@noop {} {\bibfield  {journal} {\bibinfo  {journal} {Astrophys. J.}\
  }\textbf {\bibinfo {volume} {574}},\ \bibinfo {pages} {861} (\bibinfo {year}
  {2002})}\BibitemShut {NoStop}%
\bibitem [{\citenamefont {Siemens}\ \emph {et~al.}(2006)\citenamefont
  {Siemens}, \citenamefont {Creighton}, \citenamefont {Maor}, \citenamefont
  {Majumder}, \citenamefont {Cannon},\ and\ \citenamefont {Read}}]{siemens}%
  \BibitemOpen
  \bibfield  {author} {\bibinfo {author} {\bibnamefont {Siemens}, \bibfnamefont
  {X.}}, \bibinfo {author} {\bibnamefont {Creighton}, \bibfnamefont {J.}},
  \bibinfo {author} {\bibnamefont {Maor}, \bibfnamefont {I.}}, \bibinfo
  {author} {\bibnamefont {Majumder}, \bibfnamefont {S.~R.}}, \bibinfo {author}
  {\bibnamefont {Cannon}, \bibfnamefont {K.}}, \ and\ \bibinfo {author}
  {\bibnamefont {Read}, \bibfnamefont {J.}},\ }\href@noop {} {\bibfield
  {journal} {\bibinfo  {journal} {Phys. Rev. D}\ }\textbf {\bibinfo {volume}
  {73}},\ \bibinfo {pages} {105001} (\bibinfo {year} {2006})}\BibitemShut
  {NoStop}%
\bibitem [{\citenamefont {Somiya}\ \emph {et~al.}(2012)\citenamefont {Somiya}
  \emph {et~al.}}]{Kagra}%
  \BibitemOpen
  \bibfield  {author} {\bibinfo {author} {\bibnamefont {Somiya}, \bibfnamefont
  {K.}} \emph {et~al.},\ }\href@noop {} {\bibfield  {journal} {\bibinfo
  {journal} {Class. Quant. Grav.}\ }\textbf {\bibinfo {volume} {29}},\ \bibinfo
  {pages} {124007} (\bibinfo {year} {2012})}\BibitemShut {NoStop}%
\bibitem [{\citenamefont {Spitler}\ \emph {et~al.}(2014)\citenamefont {Spitler}
  \emph {et~al.}}]{alfafrb}%
  \BibitemOpen
  \bibfield  {author} {\bibinfo {author} {\bibnamefont {Spitler}, \bibfnamefont
  {L.~G.}} \emph {et~al.},\ }\href@noop {} {\bibfield  {journal} {\bibinfo
  {journal} {Astrophys. J.}\ }\textbf {\bibinfo {volume} {790}},\ \bibinfo
  {pages} {101} (\bibinfo {year} {2014})}\BibitemShut {NoStop}%
\bibitem [{\citenamefont {Spitler}\ \emph {et~al.}(2016)\citenamefont {Spitler}
  \emph {et~al.}}]{alfarepeat}%
  \BibitemOpen
  \bibfield  {author} {\bibinfo {author} {\bibnamefont {Spitler}, \bibfnamefont
  {L.~G.}} \emph {et~al.},\ }\href@noop {} {\bibfield  {journal} {\bibinfo
  {journal} {Nature}\ }\textbf {\bibinfo {volume} {531}},\ \bibinfo {pages}
  {202} (\bibinfo {year} {2016})}\BibitemShut {NoStop}%
\bibitem [{\citenamefont {Stovall}\ \emph {et~al.}(2014)\citenamefont {Stovall}
  \emph {et~al.}}]{GBNCC}%
  \BibitemOpen
  \bibfield  {author} {\bibinfo {author} {\bibnamefont {Stovall}, \bibfnamefont
  {K.}} \emph {et~al.},\ }\href@noop {} {\bibfield  {journal} {\bibinfo
  {journal} {Astrophys. J.}\ }\textbf {\bibinfo {volume} {791}},\ \bibinfo
  {pages} {67} (\bibinfo {year} {2014})}\BibitemShut {NoStop}%
\bibitem [{\citenamefont {Sutton}\ \emph {et~al.}(2010)\citenamefont {Sutton}
  \emph {et~al.}}]{xpipeline}%
  \BibitemOpen
  \bibfield  {author} {\bibinfo {author} {\bibnamefont {Sutton}, \bibfnamefont
  {P.~J.}} \emph {et~al.},\ }\href@noop {} {\bibfield  {journal} {\bibinfo
  {journal} {New Journal of Physics}\ }\textbf {\bibinfo {volume} {12}},\
  \bibinfo {pages} {053034} (\bibinfo {year} {2010})}\BibitemShut {NoStop}%
\bibitem [{\citenamefont {Thornton}\ \emph {et~al.}(2013)\citenamefont
  {Thornton} \emph {et~al.}}]{parkesfrbs}%
  \BibitemOpen
  \bibfield  {author} {\bibinfo {author} {\bibnamefont {Thornton},
  \bibfnamefont {D.}} \emph {et~al.},\ }\href@noop {} {\bibfield  {journal}
  {\bibinfo  {journal} {Science}\ }\textbf {\bibinfo {volume} {341}},\ \bibinfo
  {pages} {53} (\bibinfo {year} {2013})}\BibitemShut {NoStop}%
\bibitem [{\citenamefont {Totani}(2013)}]{FRBBNS}%
  \BibitemOpen
  \bibfield  {author} {\bibinfo {author} {\bibnamefont {Totani}, \bibfnamefont
  {T.}},\ }\href@noop {} {\bibfield  {journal} {\bibinfo  {journal} {Pub.
  Astron. Soc. Jpn.}\ }\textbf {\bibinfo {volume} {65}},\ \bibinfo {pages}
  {L12} (\bibinfo {year} {2013})}\BibitemShut {NoStop}%
\bibitem [{\citenamefont {Trott}, \citenamefont {Tingay},\ and\ \citenamefont
  {Wayth}(2013)}]{mwafrb}%
  \BibitemOpen
  \bibfield  {author} {\bibinfo {author} {\bibnamefont {Trott}, \bibfnamefont
  {C.~M.}}, \bibinfo {author} {\bibnamefont {Tingay}, \bibfnamefont {S.~J.}}, \
  and\ \bibinfo {author} {\bibnamefont {Wayth}, \bibfnamefont {R.~B.}},\
  }\href@noop {} {\bibfield  {journal} {\bibinfo  {journal} {Astrophys. J.
  Lett.}\ }\textbf {\bibinfo {volume} {776}},\ \bibinfo {pages} {L16} (\bibinfo
  {year} {2013})}\BibitemShut {NoStop}%
\bibitem [{\citenamefont {Usov}\ and\ \citenamefont {Katz}(2000)}]{usov}%
  \BibitemOpen
  \bibfield  {author} {\bibinfo {author} {\bibnamefont {Usov}, \bibfnamefont
  {V.~V.}}\ and\ \bibinfo {author} {\bibnamefont {Katz}, \bibfnamefont
  {J.~I.}},\ }\href@noop {} {\bibfield  {journal} {\bibinfo  {journal}
  {Astronomy and Astrophysics}\ }\textbf {\bibinfo {volume} {364}},\ \bibinfo
  {pages} {655} (\bibinfo {year} {2000})}\BibitemShut {NoStop}%
\bibitem [{\citenamefont {Vachaspati}(2008)}]{vachaspati}%
  \BibitemOpen
  \bibfield  {author} {\bibinfo {author} {\bibnamefont {Vachaspati},
  \bibfnamefont {T.}},\ }\href@noop {} {\bibfield  {journal} {\bibinfo
  {journal} {Phys. Rev. Lett.}\ }\textbf {\bibinfo {volume} {101}},\ \bibinfo
  {pages} {141301} (\bibinfo {year} {2008})}\BibitemShut {NoStop}%
\bibitem [{\citenamefont {Was}(2011)}]{was}%
  \BibitemOpen
  \bibfield  {author} {\bibinfo {author} {\bibnamefont {Was}, \bibfnamefont
  {M.}},\ }\href@noop {} {\bibfield  {journal} {\bibinfo  {journal} {Ph.D.
  thesis. Universite Paris XI}\ } (\bibinfo {year} {2011})}\BibitemShut
  {NoStop}%
\bibitem [{\citenamefont {Weltevrede}, \citenamefont {Johnston},\ and\
  \citenamefont {Espinoza}(2011)}]{weltevrede}%
  \BibitemOpen
  \bibfield  {author} {\bibinfo {author} {\bibnamefont {Weltevrede},
  \bibfnamefont {P.}}, \bibinfo {author} {\bibnamefont {Johnston},
  \bibfnamefont {S.}}, \ and\ \bibinfo {author} {\bibnamefont {Espinoza},
  \bibfnamefont {C.~M.}},\ }\href@noop {} {\bibfield  {journal} {\bibinfo
  {journal} {AIP Conference Proceedings}\ }\textbf {\bibinfo {volume} {1357}},\
  \bibinfo {pages} {109} (\bibinfo {year} {2011})}\BibitemShut {NoStop}%
\bibitem [{\citenamefont {Yancey}\ \emph {et~al.}(2015)\citenamefont {Yancey}
  \emph {et~al.}}]{Yancey}%
  \BibitemOpen
  \bibfield  {author} {\bibinfo {author} {\bibnamefont {Yancey}, \bibfnamefont
  {C.}} \emph {et~al.},\ }\href@noop {} {\bibfield  {journal} {\bibinfo
  {journal} {Astrophys. J.}\ }\textbf {\bibinfo {volume} {812}},\ \bibinfo
  {pages} {168} (\bibinfo {year} {2015})}\BibitemShut {NoStop}%
\bibitem [{\citenamefont {Zink}, \citenamefont {Lasky},\ and\ \citenamefont
  {Kokkotas}(2012)}]{zink}%
  \BibitemOpen
  \bibfield  {author} {\bibinfo {author} {\bibnamefont {Zink}, \bibfnamefont
  {B.}}, \bibinfo {author} {\bibnamefont {Lasky}, \bibfnamefont {P.~D.}}, \
  and\ \bibinfo {author} {\bibnamefont {Kokkotas}, \bibfnamefont {K.~D.}},\
  }\href@noop {} {\bibfield  {journal} {\bibinfo  {journal} {Phys. Rev. D}\
  }\textbf {\bibinfo {volume} {85}},\ \bibinfo {pages} {024030} (\bibinfo
  {year} {2012})}\BibitemShut {NoStop}%
\end{thebibliography}%
\bibliographystyle{aipauth4-1}

\end{document}